\documentclass{llncs}
\usepackage{url}
\usepackage{bussproofs}
\usepackage{latexsym}
\usepackage{amsmath}
\usepackage{amssymb}
\numberwithin{equation}{section}

\newcommand{\pair}[2]{\langle #1, #2\rangle}

\newcommand{\Pred}{{\mathcal{P}}}
\newcommand{\Succ}{{\mathcal{S}}}

\newcommand{\Lg}[1]{\mathtt{#1}}

\newcommand{\ExAnd}{\wedge}
\newcommand{\ExOr}{\vee}
\newcommand{\ExNot}{\neg}

\newcommand{\ExImp}{\rightarrow}

\newcommand{\Fml}{\mathit{Fml}}
\newcommand{\Atoms}{\mathit{Atoms}}

\newcommand{\sequent}{\vdash}

\newcommand{\urule}[3]{
                                                                                        \AxiomC{#2}
                \LeftLabel{$#1$}        \UnaryInfC{#3}  
        \DisplayProof
}

\newcommand{\uruleSideCond}[4]{
        $
        \begin{array}{l}
                                                                                                \AxiomC{#2}
                        \LeftLabel{$#1$}        \UnaryInfC{#3}  
                \DisplayProof
        \\[1em]
                #4
        \end{array}
        $
}

\newcommand{\brule}[4]{
                                                                                        \AxiomC{#2}
                                                                                        \AxiomC{#3}
                \LeftLabel{$#1$}        \BinaryInfC{#4} 
        \DisplayProof
}

\newcommand{\bruleSideCond}[5]{
        $
        \begin{array}{l}
                                                                                                \AxiomC{#2}
                                                                                                \AxiomC{#3}
                        \LeftLabel{$#1$}        \BinaryInfC{#4} 
                \DisplayProof
        \\
                #5
        \end{array}
        $
}

\newcommand{\bruleSideCondEx}[5]{
        $
        \begin{array}{l}
                                                                                                \AxiomC{#2}
                                                                                                \AxiomC{#3}
                        \dashedLine \LeftLabel{$#1$}    \BinaryInfC{#4} 
                \DisplayProof
        \\
                #5
        \end{array}
        $
}

\newcommand{\Imp}{\rightarrow}
\newcommand{\Force}{\vDash}

\newcommand{\WeakImpDOne}{-\hspace{-0.05cm}<}
\newcommand{\WeakImp}{
   \begin{picture}(10,10)
     \put(1,0){$-$}
     \put(7,0){$<$}
   \end{picture}
   \;\; 
}
\newcommand{\WeakNot}{\sim \hspace{-0.1cm}}
\newcommand{\NotForce}{\nvDash}
\def\Reject{=\joinrel\mathrel|}

\newcommand{\MetaImp}{\Rightarrow}

\newcommand{\Bottom}{\perp}
\newcommand{\Top}{\top}

\newcommand{\stacked}[2]{
\genfrac{}{}{0pt}{}{#1}{#2} 
 }

\newcommand{\mycal}[1]{
        {\cal{#1}}
}

\newcommand{\Sequent}[4]{
        \left. \left. \begin{array}{l} \hspace{-5pt} \stacked{#1}{#2}  \hspace{-5pt}\end{array} \ \right|\right| #3 \sequent #4
}

\newcommand{\SequentAny}{
        \Sequent{\Succ}{\Pred}{\Gamma}{\Delta}
}

\newcommand{\Model}{
        \mycal{M}=\langle \mycal{W}, \mycal{R}, \vartheta \rangle
}

\newcommand {\BiInt}{
        \Lg{BiInt}
}

\newcommand {\GHPC}{
        \Lg{GHPC}
}

\newcommand {\Int}{
        \Lg{Int}
}

\newcommand {\DualInt}{
        \Lg{DualInt}
}

\newcommand {\SFour}{
        \Lg{S4}
}

\newcommand {\SFive}{
        \Lg{S5}
}

\newcommand {\KtSFour}{
        \Lg{Kt.S4}
}

\newcommand{\GBiInt}{\mathbf{GBiInt}}

\newcommand{\tree}[1]{
        {\mycal{#1}}
}

\newcommand{\true}{\mbox{true}}
\newcommand{\false}{\mbox{false}}

\newcommand{\IdRule}{(Id)}
\newcommand{\FalseLeftRule}{(\Bottom_L)}
\newcommand{\TrueRightRule}{(\Top_R)}
\newcommand{\AndLeftRule}{(\ExAnd_L)}
\newcommand{\AndRightRule}{(\ExAnd_R)}
\newcommand{\OrRightRule}{(\ExOr_R)}
\newcommand{\OrLeftRule}{(\ExOr_L)}

\newcommand{\ImpRightRule}{({\Imp_R})}
\newcommand{\ImpRightRuleI}{({\Imp_R^I})}

\newcommand{\ImpLeftAllRule}{(\Imp_{L})}

\newcommand{\WeakImpLeftRule}{({\WeakImp_L})}
\newcommand{\WeakImpLeftRuleI}{({\WeakImp_L^I})}

\newcommand{\WeakImpRightAllRule}{(\WeakImp_{R})}

\newcommand{\SpecialRightRule}{({\bigwedge_R})}
\newcommand{\SpecialLeftRule}{({\bigvee_L})}
\newcommand{\ReturnRule}{(Ret)}

\newcommand{\entails}{\hspace{-0.05cm}\Vdash_{_{\BiInt}}\hspace{-0.1cm}}

\newcommand{\varsTrans}[5]{
#1 := 
\left\{ \begin{array}{ll}
        #2                                                                              & \mbox{if }    #5 = \epsilon \\                                                                                                                                                                
        #3                                                                              & \mbox{if}     \text{ right prem created} \\ 
        #4                                                                              & \mbox{otherwise} \\
        \end{array}\right.
}

\newcommand{\RuleDefReturn}
{
        \uruleSideCond{\ReturnRule}
        {$$}
        {$\Sequent{\Succ := \{ \Gamma\}}{\Pred := \{ \Delta \}}{\Gamma}{\Delta}$}
        {\ \ \ \ \ \ \ \ \text{where no other rule is applicable}}
}

\newcommand{\RuleDefId}
{
        \urule{\IdRule}
        {}
        {$\Sequent{\Succ :=  \epsilon }{\Pred := \epsilon }{\Gamma, \varphi}{\Delta, \varphi}$}
}

\newcommand{\RuleDefFalseLeft}
{
        \urule{\FalseLeftRule}
        {}
        {$\Sequent{\Succ :=  \epsilon }{\Pred := \epsilon }{\Gamma, \Bottom}{\Delta}$}
}

\newcommand{\RuleDefTrueRight}
{
        \urule{\TrueRightRule}
        {}
        {$\Sequent{\Succ :=  \epsilon }{\Pred := \epsilon }{\Gamma}{\Delta, \Top}$}
}

\newcommand{\RuleDefImpLeftAll}
{
        \bruleSideCond{\ImpLeftAllRule}
        {$\Sequent{\Succ_1}{\Pred_1}{\Gamma, \varphi \Imp \psi}{\varphi, \Delta}$}
        {$\Sequent{\Succ_2}{\Pred_2}{\Gamma, \varphi \Imp \psi, \psi}{\Delta}$}
        {$\Sequent{\Succ := \Succ_1 \cup \Succ_2}{\Pred := \Pred_1 \cup \Pred_2}{\Gamma, \varphi \Imp \psi}{\Delta}$}
        {}
}

\newcommand{\RuleDefExclRightAll}
{
        \bruleSideCond{\WeakImpRightAllRule}
        {$\Sequent{\Succ_1}{\Pred_1}{\Gamma, \psi}{\Delta, \varphi \WeakImp \psi}$}
        {$\Sequent{\Succ_2}{\Pred_2}{\Gamma}{\Delta, \varphi \WeakImp \psi, \varphi}$}
        {$\Sequent{\Succ := \Succ_1 \cup \Succ_2}{\Pred := \Pred_1 \cup \Pred_2}{\Gamma}{\Delta, \varphi \WeakImp \psi}$}
        {}
}

\newcommand{\RuleDefImpRightI}
{
        \urule{\ImpRightRuleI}
        {$\Sequent{\Succ_1}{\Pred_1}{\Gamma}{\Delta, \varphi \Imp \psi, \psi}$}
        {$\Sequent{\Succ := \Succ_1}{\Pred := \Pred_1}{\Gamma}{\Delta, \varphi \Imp \psi}$}
}

\newcommand{\RuleDefExclLeftI}
{
        \urule{\WeakImpLeftRuleI}
        {$\Sequent{\Succ_1}{\Pred_1}{\Gamma, \varphi, \varphi \WeakImp \psi}{\Delta}$}
        {$\Sequent{\Succ := \Succ_1}{\Pred := \Pred_1}{\Gamma, \varphi \WeakImp \psi}{\Delta}$}
}

\newcommand{\RuleDefImpRight}
{
        \bruleSideCondEx{\ImpRightRule}
        {$\Sequent{\Succ_1}{\Pred_1}{\Gamma, \varphi}{\psi}$}
        {$\Sequent{\Succ_2}{\Pred_2}{\Gamma}{\Delta, \varphi \Imp \psi, \bigwedge \Pred_1}$}    
{$\Sequent{\varsTrans{\Succ/\Pred}{\Succ_1/\Pred_1}{\Succ_2/\Pred_2}{\{ \Gamma \}/\{ \Delta, \varphi \ExImp \psi \}}{\Pred_1}}{}{\Gamma}{\Delta, \varphi \ExImp \psi}$}
{\text{\ \ \ \ \ \ \ \ \ right prem created only if } \Pred_1 \ne \epsilon \And \forall \Pi_i \in \Pred_1 . \Pi_i \not\subseteq \{ \Delta, \varphi \ExImp \psi \}}
}

\newcommand{\RuleDefSpecialRight}
{
        \urule{\SpecialRightRule}
        {$\Sequent{\Succ_1}{\Pred_1}{\Gamma}{\Delta, \Pi_1} \ \cdots \ \Sequent{\Succ_n}{\Pred_n}{\Gamma}{\Delta, \Pi_n}$}
        {$\Sequent{\Succ := \bigcup_1^n \Succ_i}{\Pred := \bigcup_1^n \Pred_i}{\Gamma}{\Delta, \bigwedge \Pi}$}
}

\newcommand{\RuleDefExclLeft}
{
        \bruleSideCondEx{\WeakImpLeftRule}
        {$\Sequent{\Succ_1}{\Pred_1}{\varphi}{\Delta, \psi}$}
        {$\Sequent{\Succ_2}{\Pred_2}{\Gamma, \varphi \WeakImp \psi, \bigvee \Succ_1}{\Delta}$}  
        {$\Sequent{\varsTrans{\Succ/\Pred}{\Succ_1/\Pred_1}{\Succ_2/\Pred_2}{\{ \Gamma, \varphi \WeakImp \psi \}/\{ \Delta \}}{\Succ_1}}{}{\Gamma, \varphi \WeakImp \psi}{\Delta}$}
{\text{\ \ \ \ \ \ \ \ \ \ \ \ right prem created only if } \Succ_1 \ne \epsilon \And \forall \Sigma_i \in \Succ_1 . \Sigma_i \not\subseteq \{ \Gamma, \varphi \WeakImp \psi \}}        
}

\newcommand{\RuleDefSpecialLeft}
{
        \urule{\SpecialLeftRule}
        {$\Sequent{\Succ_1}{\Pred_1}{\Gamma, \Sigma_1}{\Delta} \ \cdots \ \Sequent{\Succ_n}{\Pred_n}{\Gamma, \Sigma_n}{\Delta}$}
        {$\Sequent{\Succ := \bigcup_1^n \Succ_i}{\Pred := \bigcup_1^n \Pred_i}{\Gamma, \bigvee \Sigma}{\Delta}$}
}

\newcommand{\RuleDefAndRightBlocked}
{
        \brule{\AndRightRule}
        {$\Sequent{\Succ_1}{\Pred_1}{\Gamma}{\Delta, \varphi \ExAnd \psi, \varphi}$}
        {$\Sequent{\Succ_2}{\Pred_2}{\Gamma}{\Delta, \varphi \ExAnd \psi, \psi}$}
        {$\Sequent{\Succ := \Succ_1 \cup \Succ_2}{\Pred := \Pred_1 \cup \Pred_2}{\Gamma}{\Delta, \varphi \ExAnd \psi}$}
}

\newcommand{\RuleDefAndLeftBlocked}
{
        \urule{\AndLeftRule}
        {$\Sequent{\Succ_1}{\Pred_1}{\Gamma, \varphi \ExAnd \psi, \varphi, \psi}{\Delta}$}
        {$\Sequent{\Succ := \Succ_1}{\Pred := \Pred_1}{\Gamma, \varphi \ExAnd \psi}{\Delta}$}
}

\newcommand{\RuleDefOrRightBlocked}
{
        \urule{\OrRightRule}
        {$\Sequent{\Succ_1}{\Pred_1}{\Gamma}{\Delta, \varphi \ExOr \psi, \varphi, \psi}$}
        {$\Sequent{\Succ := \Succ_1}{\Pred := \Pred_1}{\Gamma}{\Delta, \varphi \ExOr \psi}$}
}

\newcommand{\RuleDefOrLeftBlocked}
{
        \brule{\OrLeftRule}
        {$\Sequent{\Succ_1}{\Pred_1}{\Gamma, \varphi \ExOr \psi, \varphi}{\Delta}$}
        {$\Sequent{\Succ_2}{\Pred_2}{\Gamma, \varphi \ExOr \psi, \psi}{\Delta}$}
        {$\Sequent{\Succ := \Succ_1 \cup \Succ_2}{\Pred := \Pred_1 \cup \Pred_2}{\Gamma, \varphi \ExOr \psi}{\Delta}$}
}

\begin{document}
\title{A Cut-free Sequent Calculus for\\  Bi-Intuitionistic Logic: \\ Extended Version}
\author{Linda Buisman \and Rajeev Gor\'{e}}
\institute{
   The Australian National University\\
   Canberra ACT 0200, Australia
   \and
   Logic and Computation Programme\\
   Canberra Research Laboratory, NICTA\thanks{National ICT Australia is
    funded by the Australian Government's Dept of Communications,
    Information Technology and the Arts and the Australian Research
    Council through Backing Australia's Ability and the ICT Centre of
    Excellence program.}, Australia\\
   \email{\{Linda.Buisman|Rajeev.Gore\}@anu.edu.au}
}
\maketitle

\begin{abstract}
  Bi-intuitionistic logic is the extension of intuitionistic logic
  with a connective dual to implication. Bi-intuitionistic logic was
  introduced by Rauszer as a Hilbert calculus with algebraic and
  Kripke semantics. But her subsequent ``cut-free'' sequent calculus
  for $\BiInt$ has recently been shown by Uustalu to fail
  cut-elimination.  We present a new cut-free sequent calculus for
  $\BiInt$, and prove it sound and complete with respect to its Kripke
  semantics. Ensuring completeness is complicated by the interaction between
  implication and its dual, similarly to future and past modalities in
  tense logic. Our calculus handles this interaction using extended sequents
  which pass information from premises to conclusions using variables
  instantiated at the leaves of failed derivation trees.
  Our simple termination argument allows our calculus to be used for
  automated deduction, although this is not its main purpose.
\end{abstract}

\section{Introduction}

Propositional intuitionistic logic ($\Int$) has connectives $\ExImp$, $\ExAnd$, $\ExOr$ and $\ExNot$, with $\ExNot \varphi$ often defined as $\ExNot \varphi := \varphi \ExImp \Bottom$. $\Int$ has a well-known Kripke semantics, where a possible world $w$ makes $\varphi \ExImp \psi$ true if every successor $v$ that makes $\varphi$ true also makes $\psi$ true. $\Int$ also has an algebraic semantics in terms of Heyting algebras, and there is a well-known embedding from $\Int$ into the classical modal logic $\SFour$. $\Int$ is constructive in that it rejects the Law of Excluded Middle: that is, $\varphi \ExOr \ExNot \varphi$ is not a theorem of $\Int$.

Propositional dual intuitionistic logic ($\DualInt$) has connectives $\WeakImp$, $\ExAnd$, $\ExOr$ and $\WeakNot \ $, with $\WeakNot \varphi$ often defined as $\WeakNot \varphi := \Top \WeakImp \varphi$. $\DualInt$ also has Kripke semantics, where a possible world $w$ makes $\varphi \WeakImp \psi$ true if there exists a predecessor $v$ where $\varphi$ holds, but $\psi$ does not hold: that is, $\varphi$ \textit{excludes} $\psi$. Thus, the $\WeakImp$ connective of $\DualInt$ is dual to implication in $\Int$. $\DualInt$ also has algebraic semantics in terms of Brouwer algebras \cite{mckinsey1946}. There is a less well-known embedding from $\DualInt$ into $\SFour$. $\DualInt$ is para-consistent in that it rejects the Law of Non-contradiction: that is, $\varphi \ExAnd \WeakNot \varphi$ is $\DualInt$-satisfiable. Various names have been used for $\WeakImp$: coimplication \cite{wolter98,uustalu2006a}, subtraction \cite{crolard2001,crolard2004}, pseudo-difference \cite{rauszer1980}, explication \cite{rauszer1974}. We refer to it as exclusion.

Bi-intuitionistic logic ($\BiInt$), also known as subtractive logic and Heyting-Brouwer logic, is the union of $\Int$ and $\DualInt$, and it is a conservative extension of both. $\BiInt$ was first studied by Rauszer \cite{rauszer1974,rauszer1980}.
$\BiInt$ is an interesting logic to study, since it combines the constructive aspects of $\Int$ with the para-consistency of $\DualInt$.
While every $\Int$-theorem is also a $\BiInt$-theorem, adding $\DualInt$ connectives introduces a non-constructive aspect to the logic -- the disjunction property does not hold for $\BiInt$ formulae if they contain $\WeakImp$. Note that $\BiInt$ differs from intuitionistic logic with constructive negation, also known as constructible falsity \cite{nelson1949}, where the disjunction property does hold.

While the proof theory of $\Int$ and $\DualInt$ separately has been studied extensively and there are many cut-free sequent systems for $\Int$ (for example, \cite{gentzen1935,dyckhoff1992,dragalin1988}) and $\DualInt$ (for example, \cite{urbas1996,czermak1977}), the case for $\BiInt$ is less satisfactory. Although Rauszer presented a sequent calculus for $\BiInt$ in \cite{rauszer1974} and ``proved'' it cut-free, Uustalu has recently given a counter-example \cite{uustalu2004} to her cut-elimination theorem: the formula $p \ExImp (q \ExOr (r \ExImp ((p \WeakImp q) \ExAnd r))$ is $\BiInt$-valid, but cannot be derived in Rauszer's calculus without the cut rule. Similarly, Uustalu's counterexample shows that Crolard's sequent calculus \cite{crolard2001} for $\BiInt$ is not cut-free. Uustalu's counterexample fails in both Rauszer's and Crolard's calculi because they limit certain sequent rules to singleton succedents or antecedents in the conclusion, and the rules do not capture the interaction between implication and exclusion.

Uustalu and Pinto have also given a cut-free sequent-calculus for $\BiInt$ in \cite{uustalu2006a}. Since only the abstract of this work has been published so far, we have not been able to examine their sequent rules, or verify their proofs. According to the abstract \cite{uustalu2006a} and personal communication with Uustalu \cite{uustalu2006}, his calculus uses labelled formulae, thereby utilising some semantic aspects, such as explicit worlds and accessibility, directly in the rules. Hence a traditional cut-free sequent calculus for $\BiInt$ is still an open problem.

We present a new purely syntactic cut-free sequent calculus for
$\BiInt$. We avoid
Rauszer's and Crolard's restrictions on the 
antecedents and succedents for certain rules by basing our rules on
Dragalin's $\GHPC$ \cite{dragalin1988} which allows multiple formulae
on both sides of sequents. To maintain intuitionistic soundness, we
restrict 
the \emph{premise} of the implication-right
rule
to a singleton in the succedent. Dually, the
premise of our exclusion-left rule is restricted to a singleton in the
antecedent.
But using Dragalin's calculus and its dual does not give us
$\BiInt$ completeness. We therefore follow Schwendimann
\cite{schwendimann98}, and use sequents which pass relevant
information from premises to conclusions using variables instantiated
at the leaves of failed derivation trees. We then recompute parts of our derivation trees using the new information, similarly to the restart technique of \cite{horrocks1998}. Our calculus thus uses a
purely syntactic addition to traditional sequents, rather than resorting to a semantic
mechanism such as labels. Our termination argument also relies on two
new rules from \'{S}vejdar \cite{svejdar2006}.

If we were interested only in decision procedures, we could obtain a decision procedure for $\BiInt$ by embedding it
into the tense logic $\KtSFour$ \cite{wolter98}, and using tableaux for description logics with inverse roles 
\cite{horrocks1998}. However, an embedding into $\KtSFour$ provides no proof-theoretic insights into $\BiInt$ itself. Moreover, the restart technique of Horrocks et al. \cite{horrocks1998} involves
non-deterministic expansion of disjunctions, which is complicated by
inverse roles. Their actual implementation avoids this non-determinism by keeping a global view of the whole counter-model under construction. In contrast, we handle this non-determinism by syntactically encoding it using variables and extended formulae, neither of which have a semantic content. Our purely syntactic approach is preferable for proof-theoretic reasons, since models are never explicitly involved in the proof system: see Remark~\ref{proofVsCountermodel}.

The rest of the paper is organized as follows. In Section~\ref{syntaxSemantics}, we define the syntax and semantics
of $\BiInt$.  In Section~\ref{ourCalculus}, we introduce our sequent calculus $\GBiInt$ and give an example derivation of Uustalu's interaction formula. We prove the soundness and completeness of $\GBiInt$ in Sections~\ref{soundness}
and~\ref{completeness} respectively.  In Section \ref{sec:conclusion}, we outline further work.

\section{Syntax and Semantics of $\BiInt$}\label{syntaxSemantics}

In this section we introduce the syntax and semantics of $\BiInt$.

\begin{definition}[Syntax]
The formulae of $\BiInt$ are defined as:
\begin{eqnarray}
  p & ::= & \Top \mid \ \Bottom \ \mid p_0 \mid p_1 \mid \cdots  \\
  \varphi & ::= & p \mid \ExNot \varphi 
               \mid \varphi \ExAnd \varphi 
               \mid \varphi \ExOr \varphi 
               \mid \varphi \ExImp \varphi 
               \mid \varphi \WeakImp \varphi 
               \mid \ExNot \varphi
               \mid \WeakNot \varphi
\end{eqnarray}
We refer to the set of atoms as $\Atoms$, and we refer to the set of $\BiInt$ formulae as $\Fml$.
\end{definition}

The connectives $\ExNot$ and $\ExImp$ are those of intuitionistic logic, and the connectives $\WeakNot$ and $\WeakImp$ are those of dual intuitionistic logic. The connectives $\ExOr$ and $\ExAnd$ are from both.

\begin{definition}[Length]
The length of a $\BiInt$ formula $\chi$ is defined as:
$$
len(\chi) = 
\left\{
\begin{array}{lcl}
	1																				&	\ \text{ if } \ & \chi \in \Atoms \\
	len(\varphi) + 1											&	\ \text{ if } \ & \chi \in \{ \ExNot \varphi, \WeakNot \varphi \} \\
	len(\varphi) + len(\psi) + 1			&	\ \text{ if } \ & \chi \in \{ \varphi \ExOr \psi, \varphi \ExAnd \psi, \varphi \ExImp \psi, \varphi \WeakImp \psi \} .
\\
\end{array}
\right.		
$$
\end{definition}

We use the language of classical first-order logic when reasoning about $\BiInt$ at the meta-level.

\begin{definition}[Frame]
A $\BiInt$ frame is a pair $\pair{\mycal{W}}{\mycal{R}}$, where:
	\begin{enumerate}
		\item $\mycal{W}$ is a non-empty set of worlds;
		\item $\mycal{R} \subseteq \mycal{W} \times \mycal{W}$ is the binary accessibility relation;
		\item $\mycal{R}$ is reflexive, i.e.,  $\forall u \in \mycal{W} . u \mycal{R} u$;
		\item $\mycal{R}$ is transitive, i.e.,  $\forall u, v, w \in \mycal{W} . (u \mycal{R} v \And v \mycal{R} w \MetaImp u \mycal{R} w)$.
	\end{enumerate}
\end{definition}

\begin{definition}[Model]\label{model}
A $\BiInt$ model is a triple $\Model$, where:
	\begin{enumerate}
		\item $\pair{\mycal{W}}{\mycal{R}}$ is a $\BiInt$ frame;
		\item The truth valuation $\vartheta$ is a function $\mycal{W} \times \Atoms \rightarrow \{ \true, \false \}$, which tells us the truth value of an atom at a world;
		\item The persistence property holds: \\
			$\forall u,w \in \mycal{W} . \forall p \in \Atoms . (\vartheta(w,p) = \true \And w \mycal{R} u) \MetaImp (\vartheta(u,p) = \true)$;
		\item\label{top} $\forall w \in \mycal{W}. \vartheta(w,\Top) = \true$;
		\item\label{bottom} $\forall w \in \mycal{W}. \vartheta(w,\Bottom) = \false$.
	\end{enumerate}
\end{definition}

\begin{definition}[Forcing of atoms]
Given a model $\Model$, a world $w \in \mycal{W}$ and an atom $p \in \Atoms$, we write $w \Force p$ if $\vartheta(w,p) = \true$.
We pronounce $\Force$ as ``forces'', and we pronounce $\NotForce$ as ``rejects''.
\end{definition}

\begin{definition}[Forcing of formulae]\label{forcing}
Given a model $\Model$, a world $w \in \mycal{W}$ and formulae $\varphi, \psi \in \Fml$, we write:
$$
\begin{array}{lll}
	w \Force \varphi \ExOr \psi			&		\ \ \text{if}	\ \ &		w \Force \varphi \text{ or } w \Force \psi \\
	w \Force \varphi \ExAnd \psi		&		\ \ \text{if}	\ \ &		w \Force \varphi \And w \Force \psi \\
	w \Force \ExNot \varphi					&		\ \ \text{if}	\ \ &		\forall u \in \mycal{W} . [w \mycal{R} u \MetaImp (u \NotForce \varphi)] \\
	w \Force \varphi \ExImp \psi		&		\ \ \text{if}	\ \ &		\forall u \in \mycal{W} . [w \mycal{R} u \MetaImp (u \NotForce \varphi \text{ or } u \Force \psi)] \\
	w \Force \; \WeakNot \varphi		&		\ \ \text{if}	\ \ &		\exists u \in \mycal{W} . [u \mycal{R} w \And u \NotForce \varphi] \\
	w \Force \varphi \WeakImp \psi	&		\ \ \text{if}	\ \ &		\exists u \in \mycal{W} . [u \mycal{R} w \And u \Force \varphi \And u \NotForce \psi] \\
\end{array}
$$
\end{definition}

From the semantics, it can be seen that the connectives $\ExNot$ and $\WeakNot \ $ can be derived from $\ExImp$ and $\WeakImp$ respectively. Therefore from now on we restrict our attention to the connectives $\ExImp$, $\WeakImp$, $\ExAnd$, $\ExOr$ only.

\begin{lemma}
The persistence property also holds for formulae, that is:
$$\forall \Model . \forall u,w \in \mycal{W} . \forall \varphi \in \Fml . (w \Force \varphi \And w \mycal{R} u \MetaImp u \Force \varphi).$$
\end{lemma}
\begin{proof}
By induction on the length of $\varphi$. 
\end{proof}

\begin{lemma}
The reverse persistence property holds:
$$\forall \Model . \forall u,w \in \mycal{W} . \forall \varphi \in \Fml . (w \NotForce \varphi \And u \mycal{R} w \MetaImp u \NotForce \varphi).$$
\end{lemma}
\begin{proof}
Reverse persistence follows from persistence, because the truth valuation is binary. That is, suppose for a contradiction that $$\exists \Model, \exists u,w \in \mycal{W} . \exists \varphi \in \Fml . (w \NotForce \varphi \And u \mycal{R} w \And u \Force \varphi).$$
Then $u \Force \varphi$ and $u \mycal{R} w$ together with the persistence property give us $w \Force \varphi$, which contradicts $w \NotForce \varphi$.
\end{proof}

We write $\epsilon$ to mean the empty set. Given two sets of formulae $\Delta$ and $\Gamma$, we write $\Delta , \Gamma$ for $\Delta \cup \Gamma$. Given a set of formulae $\Delta$ and a formula $\varphi$, we write $\Delta , \varphi$ for $\Delta \cup \{ \varphi \}$.

\begin{definition}
Given a model $\Model$, a world $w \in \mycal{W}$ and sets of formulae $\Gamma$ and $\Delta$, we write:
$$
\begin{array}{lll}
	w \Force \Gamma			&		\text{ if }	&		\forall \varphi \in \Gamma . w \Force \varphi \\
	w \Reject \Delta	&		\text{ if }	&		\forall \varphi \in \Delta . w \NotForce \varphi. \\
\end{array}
$$
As a corollary, for any world $w$, we vacuously have $w \Force \epsilon$ and $w \Reject \epsilon$.
\end{definition}

\begin{definition}[Consequence]\label{consequence}
Given two sets $\Gamma$ and $\Delta$ of formulae, $\Gamma \entails \Delta$ means:
$$\forall \Model . \forall w \in \mycal{W} . \text{ if } w \Force \Gamma \text { then } \exists \varphi \in \Delta . w \Force \varphi.$$
We write $\Gamma \not\entails \Delta$ to mean that it is not the case that $\Gamma \entails \Delta$, that is:
$$ \exists \Model . \exists w \in \mycal{W} . (w \Force \Gamma \And w \Reject \Delta).$$
Thus $\Gamma \not\entails \Delta$ means that $\Gamma \entails \Delta$ is falsifiable.
\end{definition}

We wish to prove $\Gamma \entails \Delta$ by failing to falsify $\Gamma \entails \Delta$. By Definition~\ref{consequence}, $\Gamma \not\entails \Delta$ means that there exists a $\BiInt$ model $\Model$ that contains a world $w_0 \in \mycal{W}$ such that $w_0 \Force \Gamma$ and $w_0 \Reject \Delta$. We therefore try to construct the model using a standard \textit{counter-model construction} approach: see \cite{gallier1986}. We shall start with an initial world $w_0$ and assume that $w_0 \Force \Gamma$ and $w_0 \Reject \Delta$, and then systematically decompose the formulae in $\Gamma$ and $\Delta$. The procedure will either:
	\begin{itemize}
		\item lead to a contradiction and therefore conclude that it cannot be the case that $w_0 \Force \Gamma$ and $w_0 \Reject \Delta$, therefore $\Gamma \entails \Delta$ holds, OR
		\item construct the counter-model successfully and therefore demonstrate that it is possible that $w_0 \Force \Gamma$ and $w_0 \Reject \Delta$, therefore $\Gamma \entails \Delta$ does not hold.
	\end{itemize}

\section{Our Sequent Calculus $\GBiInt$}\label{ourCalculus}

We now present $\GBiInt$, a Gentzen-style sequent calculus for $\BiInt$. The sequents have a non-traditional component in the form of variables that are instantiated at the leaves of the derivation tree, and passed back to lower sequents from premises to conclusion. Note that the variables are not names for Kripke models and have no semantic content.

\subsection{Sequents}

First, we introduce an extended syntax that will help us in the presentation of some of our sequent rules.

\begin{definition}[Extended Syntax]\label{extendedSyntax}
The extended $\BiInt$ formulae are defined as follows:
\begin{enumerate}
	\item If $\varphi$ is a $\BiInt$ formula, then $\varphi$ is an extended $\BiInt$ formula,
	\item If $\Succ$ and $\Pred$ are sets of sets of $\BiInt$ formulae, then $\bigvee \Succ$ and $\bigwedge \Pred$ are extended $\BiInt$ formulae.
\end{enumerate}
If $ \Succ = \{ \{ \varphi_0^0, \cdots, \varphi_0^n \}, \cdots, \{ \varphi_m^0, \cdots, \varphi_m^k \} \}$ and \\ $\Pred = \{ \{ \psi_0^0, \cdots, \psi_0^n \}, \cdots, \{ \psi_m^0, \cdots, \psi_m^k \} \}$, then from every extended $\BiInt$ formula we can obtain a $\BiInt$ formula as follows:
$$
\begin{array}{c}
\bigvee \Succ \equiv (\varphi_0^0 \ExAnd \cdots \ExAnd \varphi_0^n) \ExOr \cdots \ExOr (\varphi_m^0 \ExAnd \cdots \ExAnd \varphi_m^k) \\
\bigwedge \Pred \equiv (\psi_0^0 \ExOr \cdots \ExOr \psi_0^n) \ExAnd \cdots \ExAnd  (\psi_m^0 \ExOr \cdots \ExOr \psi_m^k).
\end{array}
$$
\end{definition}

From now on,  we implicitly treat extended $\BiInt$ formulae as their $\BiInt$ equivalents. The following semantics follows directly from Definition~\ref{extendedSyntax}:

\begin{definition}[Semantics of Extended Syntax]\label{extendedSemantics}
Given a $\BiInt$ model $\Model$, and a world $w_0 \in \mycal{W}$, we write:
$$
\begin{array}{ccc}
	w \Force \bigvee \Succ			&		\text{ if }	&		\exists \Gamma \in \Succ . w \Force \Gamma \\
	w \Reject \bigwedge \Pred	&		\text{ if }	&		\exists \Delta \in \Pred . w \Reject \Delta.
\end{array}
$$
We can now extend the definition of forcing and rejecting to extended $\BiInt$ formulae in the obvious way. If $\Gamma$ and $\Delta$ are sets of extended $\BiInt$ formulae viewed as their $\BiInt$ equivalents, and $\varphi$ is an extended $\BiInt$ formula viewed as its $\BiInt$ equivalent, then:
$$
\begin{array}{lll}
	w \Force \Gamma			&		\text{ if }	&		\forall \varphi \in \Gamma . w \Force \varphi \\
	w \Reject \Delta	&		\text{ if }	&		\forall \varphi \in \Delta . w \NotForce \varphi. \\
\end{array}
$$
\end{definition}

\begin{definition}[Sequent]
A $\GBiInt$ sequent is an expression of the form $$ \SequentAny $$ and consists of the following components:
	\begin{description}
		\item[Left hand side (LHS):] $\Gamma$, a set of extended $\BiInt$ formulae;
		\item[Right hand side (RHS):] $\Delta$, a set of extended $\BiInt$ formulae;
		\item[Variables:] $\Succ$, $\Pred$, each of which is a set of sets of formulae.
	\end{description}
\end{definition}

\noindent We shall sometimes use $\Gamma \sequent \Delta$ to refer to sequents, ignoring the variable values for readability. We shall only do that in cases where the values of the variables are not important to the discussion. Note that the variables do not contain extended $\BiInt$ formulae.

We now define the meaning of a sequent in terms of the counter-model under construction.

\begin{definition}[Falsifiability]\label{falsifiability}
A sequent $$ \SequentAny $$ is falsifiable [at $w_0$ in $\mycal{M}$] if and only if there exists a $\BiInt$ model $\Model$ and $\exists w_0 \in \mycal{W}$ such that $w_0 \Force \Gamma$ and $w_0 \Reject \Delta$.
\end{definition}

\begin{definition}[Variable conditions]
We say the variable conditions of a sequent $$ \gamma = \SequentAny $$ hold if and only if $\gamma$ is falsifiable at $w_0$ in some model $\Model$ and the following conditions hold:
		\begin{description}
			\item[$\Succ$-condition:] \textbf{S}uccessor condition \\
				$\exists \Sigma \in \Succ . \forall w \in \mycal{W} .w_0 \mycal{R} w \MetaImp w \Force \Sigma$
			\item[$\Pred$-condition:] \textbf{P}redecessor condition  \\
				$\exists \Pi \in \Pred . \forall w \in \mycal{W} . w \mycal{R} w_0 \MetaImp w \Reject \Pi$				
		\end{description}		
\end{definition}

\begin{lemma}\label{notFalsifiable}
A sequent $\Gamma \sequent \Delta$ is not falsifiable if and only if $\Gamma \entails \Delta$.
\end{lemma}
\begin{proof}
Applying the negation of Definition \ref{falsifiability} to $\Gamma \sequent \Delta$ gives $\Gamma \entails \Delta$.
\end{proof}

\subsection{Sequent Rules}

\begin{definition}[Sequent Rule]
A sequent rule is of one of the forms
	$$
	\begin{array}{l}
												\AxiomC{$\gamma_1\  \cdots \ \gamma_n$}
			\LeftLabel{$(name)$}	\UnaryInfC{$\gamma_0$}	
		\DisplayProof \\
		\text{side conditions }
	\end{array}
	\ \ \ \ \ \ \ \ \ \ \ \ \ \ \ 
	\begin{array}{l}
												\AxiomC{$\gamma_1\  \cdots \ \gamma_n$}
			\LeftLabel{$(name)$}	\dashedLine \UnaryInfC{$\gamma_0$}	
		\DisplayProof \\
		\text{side conditions }
	\end{array}
	$$
where $\gamma_i$, $0 \leq i \leq n$ for $n \geq 0$, are sequents. The rule consists of the following components:
	\begin{description}
		\item[Conclusion:] $\gamma_0$, written below the horizontal line;
		\item[Premise(s):] Optional, $\gamma_1, \cdots, \gamma_n$, written above the horizontal line;
		\item[Name:] Written to the left of the horizontal line;
		\item[Side conditions:] Optional, written underneath the rule;
		\item[Branching:] Universal (indicated by a solid line) or existential (indicated by a dashed line); explained shortly.							
	\end{description}
\end{definition}

To achieve completeness and termination for $\BiInt$, we combine a number of ideas from various existing systems for $\Int$, as well as use variables for updating worlds with relevant information received from successors and predecessors. Our rules can be divided into two groups: traditional (Fig.~\ref{figTraditionalRules}) and non-traditional (Fig.~\ref{figNonTraditionalRules}).

Our \textbf{traditional} rules (Fig.~\ref{figTraditionalRules}) are
based on Dragalin's $\GHPC$ \cite{dragalin1988} for $\Int$ because we
require multiple formulae in the succedents and antecedents of
sequents for completeness; we have added symmetric rules for the
$\DualInt$ connective $\WeakImp$. The main difference
is that our $\ImpLeftAllRule$ rule 
and the symmetric $\WeakImpRightAllRule$ 
carry their principal formula and all side formulae into the premises.
Our rules for $\ExAnd$ and $\ExOr$ also carry their principal formula
into their premises to assist with termination.  Note that there are
other approaches to a terminating sequent calculus for $\Int$, e.g.,
Dyckhoff's contraction-free calculi \cite{dyckhoff1992}, or history
methods by Heuerding et al.  \cite{heuerding1998} and Howe
\cite{howe1998}. These methods are less suitable when the interaction
between $\Int$ and $\DualInt$ formulae needs to be considered, since
they erase potentially relevant formulae too soon during backward
proof search. Moreover, we found it easier to prove semantic
completeness with our loop-checking method than with history-based
methods since both \cite{heuerding1998} and \cite{howe1998} prove
completeness using syntactic transformations of derivations.
Consequently, while $\GBiInt$ is sound and complete for the $\Int$
(and $\DualInt$) fragment of $\BiInt$, it is unlikely to be as
efficient on the fragment as these specific calculi.

\begin{figure}[t]
	\begin{tabular}{cc}
	\multicolumn{2}{c}{
		\scriptsize{\RuleDefId \ \ \ \RuleDefFalseLeft \ \ \ \RuleDefTrueRight}
	}
	\\[3em]	
	\scriptsize{\RuleDefAndLeftBlocked} & \scriptsize{\RuleDefAndRightBlocked} 
	\\[3em]	
	\scriptsize{\RuleDefOrRightBlocked} & \scriptsize{\RuleDefOrLeftBlocked}
	\\[3em]
	\multicolumn{2}{c}{
		\scriptsize{\RuleDefImpLeftAll}
	}
	\\[3em]
	\multicolumn{2}{c}{
		\scriptsize{\RuleDefExclRightAll}
	}
	\\[3em]
	\multicolumn{2}{c}{
		For every rule with premises $\pi_i$ and conlusion $\gamma$, apply the rule only if:
	}
	\\
	\multicolumn{2}{c}{
		$\forall \pi_i . (LHS_{\pi_i} \not\subseteq LHS_{\gamma} \text{ or } RHS_{\pi_i} \not\subseteq RHS_{\gamma})$
	}
	\end{tabular}
\caption{$\GBiInt$ rules - traditional}
\label{figTraditionalRules}
\end{figure}

\begin{figure}[t]
	\begin{tabular}{cc}
	\multicolumn{2}{c}{
		\scriptsize{\RuleDefReturn}
	}
	\\[3em]	
	\scriptsize{\RuleDefImpRightI} & \scriptsize{\RuleDefExclLeftI}
	\\[4em]
	\multicolumn{2}{c}{
		\scriptsize{\RuleDefImpRight}
	}
	\\[4em]
	\multicolumn{2}{c}{
		\scriptsize{\RuleDefExclLeft}
	}
	\\[4em]	
		\scriptsize{\RuleDefSpecialRight} & \scriptsize{\RuleDefSpecialLeft}
	\\[3em]
	\multicolumn{2}{c}{
		\begin{minipage}{\textwidth}
		For every universally branching rule with premises $\pi_i$ and conlusion $\gamma$, \\
		apply the rule only if: $\forall \pi_i . (LHS_{\pi_i} \not\subseteq LHS_{\gamma} \text{ or } RHS_{\pi_i} \not\subseteq RHS_{\gamma})$
		\end{minipage}
	}
	\\[1em]
	\multicolumn{2}{c}{
		\begin{minipage}{\textwidth}
		For every existentially branching rule with left premise $\pi$ and conlusion $\gamma$, \\
		apply the rule only if: $LHS_{\pi} \not\subseteq LHS_{\gamma} \text{ or } RHS_{\pi} \not\subseteq RHS_{\gamma}$
		\end{minipage}
	}
	\end{tabular}
\caption{$\GBiInt$ rules - non-traditional}
\label{figNonTraditionalRules}
\end{figure}

Our rules for $\ExImp$ on the right and $\WeakImp$ on the left (Fig.~\ref{figNonTraditionalRules}) are \textbf{non-traditional}. The $\ImpRightRule$ and $\WeakImpLeftRule$ rules have {\em two} premises instead of one, and they are connected by \textbf{existential branching} as indicated by the dotted horizontal line. Existential branching means that the conclusion is derivable if {\em some} premise is derivable; thus it is dual to the conventional universal branching, where the conclusion is derivable if {\em all} premises are derivable. We chose existential branching rather than two separate non-invertible rules so the left premise can communicate information via variables to the right premise. This inter-premise communication and the use of variables is crucial to proving interaction formulae of $\BiInt$, and it gives our calculus an \textbf{operational reading}.

When applying an existential branching rule during backward proof search, we first create the left premise. If the left premise is non-derivable, then it returns the variables $\Succ_1$ and $\Pred_1$. We then use these variables to create the right premise, which corresponds to the same world as the conclusion, but with updated information. Our existential branching rules work together with $\ReturnRule$, which assigns the variables at non-derivable leaves of failed derivation trees, and $\SpecialRightRule$ and $\SpecialLeftRule$, which extract the different variable choices at existential branching rules.

The conclusion of each of our rules \textbf{assigns the variables} based on the
variables returned from the premise(s), and we use the indices $i, 1, 2$ to indicate the premise from which the variable takes its value. For rules with a single premise, the variables are simply passed down from premise to
conclusion. For example, the conclusion of $\AndLeftRule$ in Fig.~\ref{figTraditionalRules} assigns $\Succ := \Succ_1$, where $\Succ_1$ is the value of the variable at the premise. However, for rules with multiple universally branching
premises, we take a union of the sets of sets corresponding to each
falsifiable premise. For example, the conclusion of $\SpecialRightRule$ in Fig.~\ref{figNonTraditionalRules} assigns $\Succ := \bigcup_1^n \Succ_i$, where $\Succ_i$ is the value of the variable at the $i$-th premise.

This way, the sets of sets stored in our variables \textbf{determinise} the return of
formulae to lower sequents -- each non-derivable premise corresponds to an open
branch, and at this point we do not know whether it will stay open
once processed in conjunction with lower sequents. Therefore, we need
to temporarily keep all open branches: see Example~\ref{exampleFalsifiable}. Then the intuition behind adding $\bigwedge \Pred$
to the right premise of $\ImpRightRule$ is that the subsequent
application of $\SpecialRightRule$ will create one or more premises,
depending on the cardinality of $\Pred$. Since $\Pred$ is a set of
sets representing all the open branches, all of the premises of
$\SpecialRightRule$ have to be derivable in order to obtain a
derivation. On the other hand, if some premises of $\SpecialRightRule$
are non-derivable (open), we form the set that consists of the union
of the variables returned by those premises, and pass the union back
to lower sequents, and so on. The premises that are derivable
contribute only $\epsilon$ and are thus ignored by the union operator.
Also, we only create the right premise of $\ImpRightRule$ if
\textbf{every} member of $\Pred$ introduces new formulae to the
current world. Otherwise, the current world already contains one of
the open branches, which would still remain open after an application
of $\SpecialRightRule$. To summarise, the sets-of-sets concept of variables is critical to the soundness of $\GBiInt$, as it allows us to remember the required choices arising further up the tree.

The \textbf{extended syntax} allows us to syntactically encode the variable choices described above. While the variables $\Succ$ and $\Pred$ are sets of sets when we pass them down the tree and combine them using set union, we use $\bigvee \Succ$ on the left and $\bigwedge \Pred$ on the right of the sequent to reflect these choices when we add $\bigvee \Succ$ or $\bigwedge \Pred$ to the right premise of an existentially branching rule. Then the $\SpecialLeftRule$ and $\SpecialRightRule$ rules break down the extended formulae $\bigvee \Succ$ and $\bigwedge \Pred$ to yield several premises, each corresponding to one variable choice. Thus the extended syntax allows us to give an intuitive syntactic representation of the variable choices.

We have also added the rule $\ImpRightRuleI$ for implication on the
right (and dually, $\WeakImpLeftRuleI$)
originally given by \'{S}vejdar \cite{svejdar2006}. Rather than
immediately creating the successor for a rejected $\varphi
\ExImp \psi$, the $\ImpRightRuleI$ rule first pre-emptively adds
$\psi$ to the right hand side of the sequent. Although \'{S}vejdar
himself does not give the semantics behind this rule, and is unable to
explain the precise role it plays in his calculus, it is very
useful in our termination proof. The rule effectively uses the reverse
persistence property -- if some successor $v$ forces $\varphi$ and
rejects $\psi$, then the current world $w$ must reject $\psi$ too, for
if $w$ forces $\psi$, then by forward persistence so does $v$, thus
giving a contradiction.

The \textbf{side condition} on each of our rules is a general \textbf{blocking} condition, where we only explore the premise(s), if they are different from the conclusion. For example, in the $\AndRightRule$ case, the blocking condition means that we apply the rule in backward proof search only if $\varphi \not\in \Delta$ and $\psi \not\in \Delta$, since otherwise some premise would be equal to the conclusion.

$\GBiInt$ also has the \textbf{subformula property}. This is obvious
for all rules, except $\ImpRightRule$ and the dual $\WeakImpLeftRule$.
For these, the right premise ``constructs'' the formulae $\bigwedge
\Pred$ and $\bigvee \Succ$. However, since $\Pred$ and $\Succ$ are
sets of sets of subformulae of the conclusion that are again extracted
by $\SpecialRightRule$ and $\SpecialLeftRule$, the right premise of
$\ImpRightRule$ and $\WeakImpLeftRule$ effectively only contains
subformulae of the conclusion.

\begin{definition}[$\GBiInt$ tree]
A $\GBiInt$ tree for a sequent $$ \SequentAny $$ is a tree rooted at $\SequentAny$, such that:
\begin{enumerate}
	\item Each child is obtained by a backwards application of a $\GBiInt$ rule, and
	\item Each leaf is an instance of a $\FalseLeftRule$, $\TrueRightRule$, $\IdRule$ or $\ReturnRule$ rule.
\end{enumerate}
  
\end{definition}

\begin{definition}\label{derivation}
	A $\GBiInt$ tree $\tree{T}$ rooted at $ \gamma =  \Sequent{\Pred}{\Succ}{\Gamma}{\Delta} $ is a derivation if:
\begin{enumerate}
	\item $\gamma$ is the conclusion of a $\FalseLeftRule$, $\TrueRightRule$ or $\IdRule$ rule application, OR,
	\item $\gamma$ is the conclusion of a \textbf{universal branching} rule application, and \textbf{all} its \textbf{premises} are derivations, OR,
	\item $\gamma$ is the conclusion of an \textbf{existential branching} rule application, and \textbf{some} \textbf{premise} is a derivation.	
\end{enumerate}

\noindent We say that $\gamma$ is derivable if there exists a derivation for $\gamma$.

\noindent We say that $\gamma$ is not derivable if $\gamma$ has no derivation.

\end{definition}

\subsection{Examples}

In the following examples, we use a simplified version of the $\AndRightRule$ rule, which discards the principal formula from the premises, merely to save horizontal space. Also, we only show non-empty variable values.

\begin{example}

The following is a derivation tree of Uustalu's counterexample, the interaction formula $p \ExImp (q \ExOr (r \ExImp ((p \WeakImp q) \ExAnd r))$, simplified to the sequent $p \sequent q, r \ExImp ((p \WeakImp q) \ExAnd r)$. We abbreviate $X := r \ExImp ((p \WeakImp q) \ExAnd r)$. The tree should be read bottom-up while ignoring the variables $\Succ$ and $\Pred$. At the leaves, the variables are assigned and transmit information down to parents and across to some siblings. The top left application of $\ReturnRule$ occurs because an application of the $\WeakImpRightAllRule$ rule to the bolded $p \WeakImp q$ is blocked, since its left premise would not be different from its conclusion.

Notice that the key to finding the contradiction is the bolded $p \WeakImp q$ formula that is passed from the left-most leaf node back to the right premise (1) of the $\ImpRightRule$ rule. Also, the $\SpecialRightRule$ rule in $(1)$ is unary in this case, since the returned $\Pred$ variable contains only one set of formulae.

{\scriptsize{\begin{prooftree}
				\AxiomC{}
				\LeftLabel{$\ReturnRule$} \UnaryInfC{$\Sequent{\Succ := \{ \{ p, r, q \} \}}{\Pred := \{ \{ \mathbf{p \WeakImpDOne q} \} \}}{p, r, q}{\mathbf{p \WeakImp q}}$}					
				
				\AxiomC{}
				\LeftLabel{$\IdRule$} \UnaryInfC{$p, r \sequent p \WeakImp q, p$}	
								
				\LeftLabel{$\WeakImpRightAllRule$} \BinaryInfC{$\Sequent{\Succ := \{ \{ p, r, q \} \}}{\Pred := \{ \{ \mathbf{p \WeakImpDOne q} \} \}}{p, r}{p \WeakImpDOne q}$}	

				\AxiomC{}
				\LeftLabel{$\IdRule$} \UnaryInfC{$p, r \sequent r$}					
			\LeftLabel{$\AndRightRule$}	\BinaryInfC{$\Sequent{\Succ := \{ \{ p, r, q \} \}}{\Pred := \{ \{ \mathbf{p \WeakImpDOne q} \} \}}{p, r}{(p \WeakImp q) \ExAnd r}$}	
		
				\AxiomC{$(1)$}
										
		\LeftLabel{$\ImpRightRule$} \dashedLine \BinaryInfC{$p \sequent q, r \ExImp ((p \WeakImp q) \ExAnd r)$}	
\end{prooftree}}}

Where $(1)$ is:

{\scriptsize{\begin{prooftree}
					\AxiomC{}			
					\LeftLabel{$\IdRule$} \UnaryInfC{$p, \mathbf{q} \sequent q, X, p \WeakImp q$}					

					\AxiomC{}			
					\LeftLabel{$\IdRule$} \UnaryInfC{$p \sequent q, X, p \WeakImp q, \mathbf{p}$}										

				\LeftLabel{$\WeakImpRightAllRule$} \BinaryInfC{$p \sequent q, X, \mathbf{p \WeakImp q}$}	
				
				\LeftLabel{$\SpecialRightRule$} \UnaryInfC{$p \sequent q, X, \bigwedge \{ \{ \mathbf{p \WeakImp q} \} \}$}		\end{prooftree}}}

\end{example}

\begin{example}\label{exampleFalsifiable}
The following example is a $\GBiInt$-tree of a falsifiable sequent, and it shows how in the case of multiple choices for the variables, a contradiction caused by one of them does not give us a derivation. We abbreviate $Y := (\Top \WeakImp p) \ExAnd (\Top \WeakImp q)$, and $X := Y \ExImp \Bottom$.

{\scriptsize{
\begin{prooftree}
					\AxiomC{}			
					\LeftLabel{$\ReturnRule$} \UnaryInfC{$\Sequent{\Succ := \{ \{ X \} \}}{\Pred := \{ \{ \Top \WeakImp p \} \}}{X}{\Bottom, \Top \WeakImp p}$}			
	
					\AxiomC{}			
					\LeftLabel{$\ReturnRule$} \UnaryInfC{$\Sequent{\Succ := \{ \{ X \} \}}{\Pred := \{ \{ \mathbf{\Top \WeakImp q} \} \}}{X}{\Bottom, \mathbf{\Top \WeakImp q}}$}			

				\LeftLabel{$\AndRightRule$} \BinaryInfC{$\Sequent{\Succ := \{ \{ X \} \}}{\Pred := \left\{ \stacked{\{ \Top \WeakImp p \},}{\{ \mathbf{\Top \WeakImp q} \}} \right\} }{X}{\Bottom, Y}$}	

				\LeftLabel{$\IdRule$} \AxiomC{$$}							
				\LeftLabel{$\FalseLeftRule$} \UnaryInfC{$X, \Bottom \sequent \Bottom$}

			\LeftLabel{$\ImpLeftAllRule$}	\BinaryInfC{$\Sequent{\Succ := \{ \{ X \} \}}{\Pred := \left\{ \stacked{\{ \Top \WeakImp p \},}{\{ \mathbf{\Top \WeakImp q} \}} \right\} }{X}{\Bottom}$}	

			\AxiomC{$(2)$}
										
		\LeftLabel{$\ImpRightRule$} \dashedLine \BinaryInfC{$\Sequent{\Succ := \{ \{ q \} \} }{\Pred := \left\{ \left\{ \stacked{ p, X \ExImp \Bottom,}{\mathbf{\Top \WeakImp q} } \right\} \right\}}{}{p, X \ExImp \Bottom}$}	
\end{prooftree}
}}

Where $(2)$ is:

{\scriptsize{\begin{prooftree}
					\AxiomC{}			
					\LeftLabel{$\IdRule$} \UnaryInfC{$\vdots$}			

					\AxiomC{}			
					\LeftLabel{$\TrueRightRule$} \UnaryInfC{$\vdots$}			

					\LeftLabel{$\WeakImpRightAllRule$} \BinaryInfC{$\sequent p, X \ExImp \Bottom, \Top \WeakImp p$}			

					\AxiomC{}
					\LeftLabel{$\ReturnRule$} \UnaryInfC{$\Sequent{\Succ := \{ \{ q \} \} }{\Pred := \left\{ \left\{ \stacked{ p, X \ExImp \Bottom,}{\mathbf{\Top \WeakImp q}} \right\} \right\}}{q}{p, X \ExImp \Bottom, \mathbf{\Top \WeakImp q}}$}			
					
					\AxiomC{}			
					\LeftLabel{$\TrueRightRule$} \UnaryInfC{$\vdots$}			
							
					\LeftLabel{$\WeakImpRightAllRule$} \BinaryInfC{$\Sequent{\Succ := \{ \{ q \} \} }{\Pred := \left\{ \left\{ \stacked{ p, X \ExImp \Bottom,}{\mathbf{\Top \WeakImp q} } \right\} \right\}}{}{p, X \ExImp \Bottom, \Top \WeakImp q}$}

				\LeftLabel{$\SpecialRightRule$} \BinaryInfC{$\Sequent{\Succ := \{ \{ q \} \} }{\Pred := \left\{ \left\{ \stacked{ p, X \ExImp \Bottom,}{\mathbf{\Top \WeakImp q} } \right\} \right\}}{}{p, X \ExImp \Bottom, \bigwedge \left\{ \{ \Top \WeakImp p \}, \{ \mathbf{\Top \WeakImp q} \} \right\}}$}		\end{prooftree}}}

In this case, the $\SpecialRightRule$ rule in $(2)$ has two premises, since the returned $\Pred$ variable contains two sets of formulae. Since only the left premise of the $\SpecialRightRule$ rule is derivable, the conclusion is not derivable. Thus, the open branch corresponding to the bolded member $\{ \Top \WeakImp q \}$ of $\Pred$ remains open. If we did not return both variable choices from the left sibling of $(2)$, then we might mistakenly derive $(2)$ without seeing this open branch.
\end{example}

\begin{lemma}
If a $\GBiInt$-tree $\tree{T}$ rooted at $ \gamma = \Sequent{\Pred}{\Succ}{\Gamma}{\Delta} $ is a derivation then $\Succ = \Pred = \epsilon$.
\end{lemma}
\begin{proof}
By induction on the longest branch in $\tree{T}$.
\end{proof}

\subsection{Termination Proof}\label{sec:termination}

We first show that proof search in $\GBiInt$ terminates because the subsequent soundness proof relies on our ability to receive the variables from the left premises of transitional rules.

\begin{definition}
The rules of $\GBiInt$ are categorised as follows:
	\begin{description}
		\item[Operational:] $\ReturnRule$;
		\item[Logical:] $ $
		\begin{description}
			\item[Static:] $\IdRule$, $\FalseLeftRule$, $\TrueRightRule$, $\AndLeftRule$, $\OrLeftRule$, $\AndRightRule$, $\OrRightRule$, $\ImpLeftAllRule$, $\WeakImpRightAllRule$, $\ImpRightRuleI$, $\WeakImpLeftRuleI$;
			\item[Transitional:] $\ImpRightRule$, $\WeakImpLeftRule$;
			\item[Special:] $\SpecialLeftRule$, $\SpecialRightRule$.			
		\end{description}
	\end{description}
\end{definition}

The intuition behind the classification of the logical rules is that the static rules add formulae to the current world in the counter-model, the transitional rules create new worlds and add formulae to them, and the special rules decompose variables returned from non-derivable leaves. We shall prove this formally for each rule later. The classification justifies the following search strategy.

\begin{figure}[t]
\textbf{Function} Prove \\
Input: sequent $\gamma_0$ \\
Output: Derivable ($true$ or $false$)
\begin{enumerate}
	\item If $\rho \in \{ \IdRule, \FalseLeftRule, \TrueRightRule\}$ applicable to $\gamma_0$ then
		\begin{enumerate}
			\item Return $true$
		\end{enumerate}
	\item Else if any special or static rule $\rho$ applicable to $\gamma_0$ then
		\begin{enumerate}
			\item Let $\gamma_1, \cdots, \gamma_n$ be the premises of $\rho$
			\item Return $\bigwedge Prove(\gamma_i)$
		\end{enumerate}
	\item Else for each transitional rule $\rho$ applicable to $\gamma_0$ do
		\begin{enumerate}
			\item Let $\gamma_1$ and $\gamma_2$ be the premises of $\rho$
			\item If $\bigvee Prove(\gamma_i) = true$ then return $true$
		\end{enumerate}
	\item Endif
	\item Return $false$.	
\end{enumerate}
\caption{Proof search strategy. Note that we have left out the variables for simplicity. $\bigwedge_{i=1}^{n} Prove(\gamma_i)$ is true iff $Prove(\gamma_i)$ is true for all premises $\gamma_i$ for $1 \leq i \leq n$, and $\bigvee_{i \in \{1, 2\}} Prove(\gamma_i)$ is true iff $Prove(\gamma_i)$ is true for some premise $\gamma_i$ for $i \in \{1, 2\}$.}
\label{strategyFig}
\end{figure}

\begin{definition}[Strategy]\label{strategy}
The strategy defined in Figure~\ref{strategyFig} is used when applying the rules of our sequent calculus in backward proof search. Note that we have left out the variables for simplicity.
\end{definition}

\begin{definition}[Subformulae]\label{sf}
For a $\BiInt$ formula, we define the subformulae as follows, where $p \in \Atoms$ and $\varphi, \psi \in \Fml$:
$$
\begin{array}{ll}
	sf(p) 								  		& = \{ p \} \\
	sf(\varphi \ExOr \psi)  		& = sf(\varphi) \cup sf(\psi) \cup \{ \varphi \ExOr \psi \} \\
	sf(\varphi \ExAnd \psi) 		& = sf(\varphi) \cup sf(\psi) \cup \{ \varphi \ExAnd \psi \} \\
	sf(\varphi \ExImp \psi)			& = sf(\varphi) \cup sf(\psi) \cup \{ \varphi \ExImp \psi \} \\
	sf(\varphi \WeakImp \psi)		& = sf(\varphi) \cup sf(\psi) \cup \{ \varphi \WeakImp \psi \} \\	
	sf(\bigvee \Succ)  					& = \displaystyle{\bigcup_{\Sigma \in \Succ}sf(\Sigma)} \\
	sf(\bigwedge \Pred) 				& = \displaystyle{\bigcup_{\Pi \in \Pred}sf(\Pred)}\\
\end{array}
$$
For a set $\Gamma$ of extended $\BiInt$ formulae, we define $sf(\Gamma) = \displaystyle{\bigcup_{\chi \in \Gamma}sf(\chi)}$.
\end{definition}

Note that the subformulae of $\bigvee \Succ$ and $\bigwedge \Pred$ do not include the conjunctions and disjunctions implicit in their $\BiInt$ equivalents.

\begin{definition}[LEN]
Let $>_{len}$ be a lexicographic ordering of sequents:
$$
\begin{array}{rcl}
	(\Gamma_2 \sequent \Delta_2) >_{len} (\Gamma_1 \sequent \Delta_1) & \text{\ \ iff \ } & |\Gamma_2| > |\Gamma_1| \text{ or } \\
	& & |\Gamma_2| = |\Gamma_1|$ and $|\Delta_2| > |\Delta_1|
\end{array}
$$
\end{definition}

\begin{definition}
Given a $\GBiInt$-tree $\tree{T}$ and a branch $\mycal{B}$ in $\tree{T}$, we say that $\mycal{B}$ is \textbf{forward-only} if $\mycal{B}$ contains only applications of static and special rules, $\ImpRightRule$ and the right premises of $\WeakImpLeftRule$. Similarly, $\mycal{B}$ is \textbf{backward-only} if $\mycal{B}$ contains only applications of static and special rules, $\WeakImpLeftRule$ and the right premises of $\ImpRightRule$. A branch is \textbf{single-directional} if it is either forward-only or backward-only. Finally, a branch contains \textbf{interleaved} left premises of transitional rules if it contains a sequence $\langle \cdots, \gamma_i, \cdots, \gamma_j, \cdots, \gamma_k, \cdots \rangle$ such that $\gamma_i$ is the left premise of $\ImpRightRule$, $\gamma_j$ is the left premise of $\WeakImpLeftRule$, and $\gamma_k$ is the left premise of $\ImpRightRule$.
\end{definition}

\begin{lemma}\label{forwardOnlyFinite}
Every forward-only branch of any $\GBiInt$-tree is finite.
\end{lemma}
\begin{proof}
We show that on every such branch, the length of a sequent defined according to $>_{len}$ increases.

Consider a rule $\rho$, and a backwards application of $\rho$ to some $\Gamma \sequent \Delta$, which yields $n$ premises $\Gamma_i \sequent \Delta_i$, where $1 \leq i \leq n$.

We show that if $\rho$ is a static rule, then for all premises $i$, we have $(\Gamma_i \sequent \Delta_i) >_{len} (\Gamma \sequent \Delta)$:
\begin{description}
	\item[\rm{$\rho \in \{ \AndLeftRule$, $\OrLeftRule$, $\WeakImpLeftRuleI \}$:}] Then $|\Gamma_i| > |\Gamma|$;
	\item[\rm{$\rho = \ImpRightRuleI$:}] Then $|\Gamma_1| = |\Gamma|$ and $|\Delta_1| > |\Delta|$;	
	\item[\rm{$\rho = \ImpLeftAllRule$:}] Then for the left premise, $|\Gamma_1| = |\Gamma|$ and $|\Delta_1| > |\Delta|$, and for the right premise, $|\Gamma_2| > |\Gamma|$;
	\item[\rm{$\rho = \WeakImpRightAllRule$:}] Then for the left premise, $|\Gamma_1| > |\Gamma|$, and for the right premise, $|\Gamma_2| = |\Gamma|$ and $|\Delta_2| > |\Delta|$.
\end{description}

We now show the cases for $\rho \in \{\ImpRightRule, \WeakImpLeftRule, \SpecialRightRule, \SpecialLeftRule \}$.
Even though the right premise of $\ImpRightRule$ and $\WeakImpLeftRule$ itself is not greater than the conclusion, we show that the lemma holds on the overall $\GBiInt$ branch, since according to the strategy we immediately apply $\SpecialRightRule$ or $\SpecialLeftRule$, thus increasing the length of the premise according to $>_{len}$.
\begin{description}
	\item[\rm{$\rho = \ImpRightRule$:}] For every $\ImpRightRule$ rule application:
		\begin{enumerate} 
			\item Consider the left premise $\Gamma_1 \sequent \Delta_1$. We know that according to our strategy, the $\ImpRightRuleI$ rule has already been applied and thus $\psi \in \Delta$, so $\ImpRightRule$ is applied only if $\varphi \not\in \Gamma$. Therefore, for the left premise, we have $|\Gamma_1| > |\Gamma|$;
			\item\label{rightPremImp} Consider the right premise $\Gamma_2 \sequent \Delta_2$. It is created only if
\begin{equation}\label{conditionForRightPrem}
	\Pred_1 \ne \epsilon \And \forall \Pi_i \in \Pred_1 . \Pi_i \not\subseteq \{ \Delta, \varphi \ExImp \psi \}.
\end{equation}
			That is, every member of $\Pred_1$ introduces new formulae to the RHS. But recall that $sf(\bigwedge \Pred_1) \subseteq sf(\Gamma \cup \Delta)$. According to our strategy, the $\SpecialRightRule$ rule will be immediately applied to $\bigwedge \Pred_1$ in $\Delta_2$, giving $n \geq 1$ premises $\Gamma_2^j \sequent \Delta_2^j$ where $1 \leq j \leq n$. By \ref{conditionForRightPrem}, we will then have $|\Delta_2^j| > |\Delta|$ for all $j$. We also have $|\Gamma_2^j| = |\Gamma|$ for all $j$. Therefore, according to the lexicographic ordering, we have $(\Gamma_2^j \sequent \Delta_2^j) >_{len} (\Gamma \sequent \Delta)$ for all the premises $\Gamma_2^j \sequent \Delta_2^j$.
		\end{enumerate}			
	\item[\rm{$\rho = \SpecialRightRule$:}] Since the $\SpecialRightRule$ rule is only used in conjunction with the right premise of the $\ImpRightRule$ rule, see case \ref{rightPremImp} above;						
	\item[\rm{$\rho = \WeakImpLeftRule$:}] For every $\WeakImpLeftRule$ rule application:
		\begin{enumerate} 
			\item The assumption of the lemma does not apply to the left premise;
			\item The case for the right premise is dual to the case for $\ImpRightRule$ above.
		\end{enumerate}		
	\item[\rm{$\rho = \SpecialLeftRule$:}] By symmetry with the case for $\SpecialRightRule$ above;			
\end{description}
Since the length of a sequent defined according to $>_{len}$ increases on every forward-only branch as shown above, and since $\GBiInt$ has the subformula property, eventually no more formulae can be added to a sequent on a forward-only branch, and the branch will terminate.
\end{proof}

\begin{lemma}\label{backwardOnlyFinite}
Every backward-only branch of any $\GBiInt$-tree is finite.
\end{lemma}
\begin{proof}
By symmetry with Lemma~\ref{forwardOnlyFinite}.
\end{proof}

\begin{lemma}\label{infMustBeInterleaved}
If a $\GBiInt$-tree contains an infinite branch, then the branch contains an infinite number of interleaved left premises of transitional rules.
\end{lemma}
\begin{proof}
By Lemmas \ref{forwardOnlyFinite} and \ref{backwardOnlyFinite}, single-directional branches must eventually terminate. Thus, a potential infinite loop must involve an infinite number of interleaved left premises of transitional rules $\ImpRightRule$ and $\WeakImpLeftRule$.
\end{proof}

\begin{definition}[Degree]
The degree of a $\BiInt$ formula $\chi$ is defined as:
$$
deg(\chi) = 
\left\{
\begin{array}{lcl}
	0																	&	\text{ if } & \chi \in \Atoms \\
	deg(\varphi) + deg(\psi)					&	\text{ if } & \chi \in \{ \varphi \ExOr \psi, \varphi \ExAnd \psi \} \\
	deg(\varphi) + deg(\psi) + 1			&	\text{ if } & \chi \in \{ \varphi \ExImp \psi, \varphi \WeakImp \psi \} \\
\end{array}
\right.		
$$

Thus, the degree of $\varphi$ is the number of $\ExImp$ and $\WeakImp$ connectives in $\varphi$.

The degree of a sequent $\Gamma \sequent \Delta$ is defined as:
$$deg(\Gamma \sequent \Delta) = \displaystyle {\sum_{\varphi \in sf(\Gamma \cup \Delta)}deg(\varphi)}$$
\end{definition}

Note that we have deliberately defined the degree of a sequent as the sum of the degrees of subformulae, because it allows us to make the following observations, which will be crucial in the main termination proof.

\begin{corollary}\label{cannotIncrease}
Since $\GBiInt$ has the subformula property, the degree of a sequent can never increase in backward proof search. In other words, no $\GBiInt$ rule can increase the degree of a sequent.
\end{corollary}

\begin{corollary}\label{mustDecrease}
Given two sequents $\gamma_1$ and $\gamma_2$, if $sf(\gamma_2) \subsetneq sf(\gamma_1)$, then $deg(\gamma_2) < deg(\gamma_1)$. That is, removing some formula $\varphi$ from a sequent during backward proof search decreases the degree of the sequent if $\varphi$ is not a subformula of any other formula in the sequent, since $\varphi$ no longer contributes to the sum of degrees of subformulae.
\end{corollary}

\begin{theorem}[Termination]\label{termination}
Every $\GBiInt$-tree constructed according to the strategy of Definition~\ref{strategy} is finite.
\end{theorem}
\begin{proof}
Suppose for a contradiction that there exists an infinite $\GBiInt$-tree $\tree{T}$. Since every rule has a finite number of premises, i.e., finite branching, then by K\"{o}nig's lemma an infinite tree can only be obtained by having a branch of infinite length. Thus, $\tree{T}$ has an infinite branch $\mycal{B}$. By Lemma~\ref{infMustBeInterleaved}, $\mycal{B}$ must contain an infinite number of interleaved left premises of transitional rules, as shown below:
\begin{prooftree}
				\AxiomC{$\vdots$}
				\UnaryInfC{$\pi_2 = (\Gamma_2, \varphi_2 \sequent \psi_2)$}											

				\AxiomC{$\vdots$}													
				\UnaryInfC{$\pi_2^r$}
		
			\LeftLabel{$\ImpRightRule$}	\dashedLine \BinaryInfC{$\Gamma_2 \sequent \Delta_2, \varphi_2 \ExImp \psi_2$}											
			\UnaryInfC{$\vdots$}
	 		\UnaryInfC{$\varphi_1 \sequent \psi_1, \Delta_1$}													 		

			\AxiomC{$\vdots$}													
			\UnaryInfC{$\pi_1^r$}

		\LeftLabel{$\WeakImpLeftRule$} \dashedLine \BinaryInfC{$\Gamma_1, \varphi_1 \WeakImp \psi_1 \sequent \Delta_1$}	

		\UnaryInfC{$\vdots$}													
		\UnaryInfC{$\Gamma_0, \varphi_0 \sequent \psi_0$}											

		\AxiomC{$\vdots$}													
		\UnaryInfC{$\pi_0^r$}
	\LeftLabel{$\ImpRightRule$} \dashedLine \BinaryInfC{$\pi_0 = (\Gamma_0 \sequent \Delta_0, \varphi_0 \ExImp \psi_0)$}
	\UnaryInfC{$\vdots$}															
\end{prooftree}

Let $\chi \in sf(\pi_0)$ be some formula such that $deg(\chi) = max(\{ deg(\varphi)\ | \ \varphi \in sf(\pi_0)\})$, that is, $\chi$ is one of the subformulae with the maximum degree. In particular, this means that $\chi$ is not a subformula of any formula with a larger degree. We shall now show that $\chi \not\in sf(\pi_2)$.

There are two cases:
\begin{description}
	\item[$\chi \not\in sf(\Gamma_0)$:]	Then $\chi \in sf(\Delta_0)$ or $\chi = \varphi_0 \ExImp \psi_0$. In both cases, $\chi \not\in sf(\pi_2)$.
	\item[$\chi \in sf(\Gamma_0)$:] Then it cannot be the case that $\chi \in sf(\varphi_1)$ or $\chi \in sf(\psi_1)$, since then $deg(\varphi_1 \WeakImp \psi_1) > deg(\chi)$, contradicting our assumption that $deg(\chi) = max(\{ deg(\varphi) \ | \ \varphi \in sf(\pi_0)\})$. Therefore, either:
	\begin{itemize}
		\item $\chi$ and all its occurrences in subformulae disappear from the sequent at the premise of $\WeakImpLeftRule$, in which case $\chi \not\in sf(\pi_2)$, or
		\item $\chi$ is moved to the RHS of the sequent by applying the $\ImpLeftAllRule$ rule to some formula $\chi \ExImp \tau$. However, since $deg(\chi \ExImp \tau) > deg(\chi)$, it again contradicts our assumption that $deg(\chi) = max(\{ deg(\varphi) \ | \ \varphi \in sf(\pi_0)\})$.	
	\end{itemize} 
\end{description}

We have shown that for some formula $\chi$ we have $\chi \in sf(\pi_0)$ and $\chi \not\in sf(\pi_2)$. Also, by the subformula property of $\GBiInt$ we have $sf(\pi_2) \subseteq sf(\pi_0)$. Together with $\chi \in sf(\pi_0)$ and $\chi \not\in sf(\pi_2)$, this means $sf(\pi_2) \subsetneq sf(\pi_0)$. Then by Corollary~\ref{mustDecrease} we have $deg(\pi_2) < deg(\pi_0)$. Note that the steps indicated by vertical ellipses ($\vdots$) are arbitrary, since by Corollary~\ref{cannotIncrease} no rule can increase the degree of a sequent.
	
Since we have $deg(\pi_2) < deg(\pi_0)$, we know that every sequence of interleaved transitional rule applications must decrease the degree of the sequent. This can only happen a finite number of times, until no more transitional rules are applicable. Therefore our assumption was wrong, and no branch $\mycal{B}$ can be infinite. Therefore, every $\GBiInt$-tree is finite.
\end{proof}

\section{Soundness}\label{soundness}

\subsection{Proof Outline}

Instead of the traditional approach of showing that each rule application preserves validity downwards, we use the notion of falsifiability and show that each rule application preserves falsifiability upwards. We then use Lemma~\ref{notFalsifiable} to make the connection between falsifiability and validity.

Also, our addition of variables to the calculus introduces a two-way flow of information in the $\GBiInt$ trees, and this complicates the usually simple soundness proof.

We separate the notion of soundness into two: \textit{local soundness}, applicable locally to a single rule application, and \textit{global soundness}, which takes into account the propagation of variables from the leaves down to some node, and possible instances of the operational $\ReturnRule$ rule. Note that locality here refers to locality in the $\GBiInt$ trees, not locality in the underlying Kripke models. We use the notions of static and transitional rules to classify the rules according to this latter notion.

\subsection{Local soundness}

\begin{definition}[Local soundness]
A logical rule in $\GBiInt$ is locally sound if and only if:
\begin{itemize}
	\item For rules with \textbf{universal} branching: if the conclusion is falsifiable, then some premise is falsifiable;
	\item For rules with \textbf{existential} branching: if the conclusion is falsifiable, then all premises are falsifiable.
\end{itemize}
\end{definition}

We shall now show that each static and special rule is locally sound, and we shall then use induction on the height of a derivation tree to extend our proof to arbitrary trees containing static rules, special rules, transitional rules and the operational $\ReturnRule$ rule.

\begin{lemma}\label{staticRuleSoundness}
Each static and special rule of $\GBiInt$ is locally sound.
\end{lemma}
\begin{proof}
We consider each static and special rule in turn. We assume that the conclusion is falsifiable, and show that some premise is falsifiable.
\begin{enumerate}
	
	\item $ $ \\ \RuleDefId \\[1em]
		The conclusion of this rule is never falsifiable, because no $\BiInt$ model can contain a world $w$ such that $w \Force \varphi$ and $w \NotForce \varphi$.
	
	\item $ $ \\ \RuleDefFalseLeft \\[1em]
		The conclusion of this rule is never falsifiable, because by Property~\ref{bottom} of Definition~\ref{model}, no $\BiInt$ model can contain a world $w$ such that $w \Force \Bottom$.
	
	\item $ $ \\ \RuleDefTrueRight \\[1em]
		The conclusion of this rule is never falsifiable, because by Property~\ref{top} of Definition~\ref{model}, no $\BiInt$ model can contain a world $w$ such that $w \NotForce \Top$.
	
	\item $ $ \\ \RuleDefAndRightBlocked \\[1em]
		Since the conclusion is falsifiable by assumption, we know from Definition \ref{falsifiability} that there exists a world $w_0$ such that:
		\begin{description}
			\item[\rm{(i)}] $w_0 \Force \Gamma$ and
			\item[\rm{(ii)}] $w_0 \Reject \Delta, \varphi \ExAnd \psi$.
		\end{description}
		From the semantics of $\ExAnd$ in $\BiInt$, (b) implies that either:
		\begin{description}
			\item[\rm{(ii.1)}] $w_0 \Reject \Delta, \varphi \ExAnd \psi, \varphi$ or
			\item[\rm{(ii.2)}] $w_0 \Reject \Delta, \varphi \ExAnd \psi, \psi$.
		\end{description}
		
		To show that some premise of the $\AndRightRule$ rule is falsifiable, we need to show that there exists a world $w'$ such that some premise is falsifiable at $w'$. We let $w' = w_0$.
		
		Then case (ii.1) together with (i) gives us that the left premise is falsifiable, or case (ii.2) together with (i) gives us that the right premise is falsifiable.
		
	\item $ $ \\ \RuleDefOrLeftBlocked \\[1em]
		By symmetry with the $\AndRightRule$ rule.

	\item $ $ \\ \RuleDefOrRightBlocked \\[1em]
		Since the conclusion is falsifiable by assumption, we know from Definition \ref{falsifiability} that there exists a world $w_0$ such that:
		\begin{description}
			\item[\rm{(i)}] $w_0 \Force \Gamma$ and
			\item[\rm{(ii)}] $w_0 \Reject \Delta, \varphi \ExOr \psi$
		\end{description}

		To show that the premise of the $\OrRightRule$ rule is falsifiable, we need to show that there exists a world $w'$ such that the premise is falsifiable at $w'$. We let $w' = w_0$.		
		
		From the semantics of $\ExOr$ in $\BiInt$, (ii) implies that $w_0 \Reject \Delta, \varphi \ExOr \psi, \varphi$ and $w_0 \Reject \Delta, \varphi \ExOr \psi, \psi$. Together with (i), this means that the premise is falsifiable.

	\item $ $ \\ \RuleDefAndLeftBlocked \\[1em]
		By symmetry with the $\OrRightRule$ rule.		
			
	\item $ $ \\ \RuleDefImpLeftAll  \\[1em]
		Since the conclusion is falsifiable by assumption, we know from Definition \ref{falsifiability} that there exists a world $w_0$ such that:
		\begin{description}
			\item[\rm{(i)}] $w_0 \Force \Gamma, \varphi \ExImp \psi$ and
			\item[\rm{(ii)}] $w_0 \Reject \Delta$.
		\end{description}
		From the semantics of $\ExImp$ in $\BiInt$, (i) implies that for all successors $w$, we have $w \NotForce \varphi$ or $w \Force \psi$.
		
		By reflexivity of $\mycal{R}$, this applies to $w_0$ too, so we have:
		\begin{description}
			\item[\rm{(i.1)}] $w_0 \NotForce \varphi$ or
			\item[\rm{(i.2)}] $w_0 \Force \psi$.
		\end{description}

	To show that some premise of the $\ImpLeftAllRule$ rule is falsifiable, we need to show that there exists a world $w'$ such that some premise is falsifiable at $w'$. We let $w' = w_0$.

		Then items (i), (ii) and (i.1) give us that the left premise is falsifiable, or items (i), (ii) and (i.2) give us that the right premise is falsifiable.

	\item $ $ \\ \RuleDefExclRightAll \\[1em]
		By symmetry with $\ImpLeftAllRule$.

	\item $ $ \\ \RuleDefImpRightI \\[1em]
		Since the conclusion is falsifiable by assumption, we know from Definition \ref{falsifiability} that there exists a world $w_0$ such that:
		\begin{description}
			\item[\rm{(i)}] $w_0 \Force \Gamma$ and
			\item[\rm{(ii)}] $w_0 \Reject \Delta, \varphi \ExImp \psi$.
		\end{description}
	From the semantics of $\ExImp$ in $\BiInt$, (ii) implies that there exists a successor $w_1$ such that:
		\begin{description}
			\item[\rm{(iii)}] $w_0 \mycal{R} w_1$ and 		
			\item[\rm{(iv)}] $w_1 \Force \varphi$ and 
			\item[\rm{(v)}] $w_1 \NotForce \psi$.
		\end{description}	
	Then, by the reverse persistence property of $\BiInt$, and (iii) and (v), we have:
		\begin{description}
			\item[(vi)] $w_0 \NotForce \psi$.
		\end{description}

		To show that the premise of the $\ImpRightRule$ rule is falsifiable, we need to show that there exists a world $w'$ such that the premise is falsifiable at $w'$. We let $w' = w_0$.
		
		Then items (i), (ii) and (vi) give us that the premise is falsifiable.
	
	\item $ $ \\ \RuleDefExclLeftI \\[1em]	
			By symmetry with $\ImpRightRuleI$.
			
	\item $ $ \\ \RuleDefSpecialLeft \\[1em]
		Since the conclusion is falsifiable by assumption, we know from Definition \ref{falsifiability} that there exists a world $w_0$ such that:
		\begin{description}
			\item[\rm{(i)}] $w_0 \Force \Gamma, \bigvee \Sigma$ and
			\item[\rm{(ii)}] $w_0 \Reject \Delta$.
		\end{description}
	From the semantics of $\bigvee \Sigma$ (recall Definition~\ref{extendedSemantics}), (i) implies that:
		\begin{description}
			\item[\rm{(iii)}] for some $\Sigma_i \in \Sigma$, we have $w_0 \Force \Sigma_i$.
		\end{description}

		To show that some premise of the $\SpecialLeftRule$ rule is falsifiable, we need to show that there exists a world $w'$ such that this premise is falsifiable at $w'$. We let $w' = w_0$.
	
		Then items (i), (ii) and (iii) give us that the $i$-th premise containing $\Sigma_i$ is falsifiable at $w_0$.
	
	\item $ $ \\ \RuleDefSpecialRight \\[1em]
		By symmetry with $\SpecialLeftRule$.

\end{enumerate}

\end{proof}

\begin{remark}
Note that the static rules also preserve falsifiability downwards: if some premise $\pi$ is falsifiable, then the conclusion $\gamma$ is falsifiable. This is easy to see, since we have $LHS_\pi \supseteq LHS_\gamma$ and $RHS_\pi \supseteq RHS_\gamma$.
\end{remark}

\subsection{Global soundness}

We have shown that all the static and special rules preserve falsifiability upwards, in other words, they are \textit{locally sound}. Since the $\Succ$ and $\Pred$ variables propagate downwards, from the leaves to the root, we can only reason about the variable conditions of rules when we consider an entire tree rooted at a rule application. Similarly, since the soundness of the transitional rules relies on the variables, we can only reason about it we consider an entire tree rooted at a transitional rule application. We shall now show that $\GBiInt$ rules are \textit{globally sound}, that is, they preserve falsifiability upwards \textit{and} variable conditions downwards.

\begin{lemma}[Global soundness]\label{globalSoundness}
Given any $\GBiInt$ tree $\tree{T}$, for every sequent $\gamma_0 \in \tree{T}$, the following holds: if $\gamma_0$ is falsifiable, then:
\begin{enumerate}
	\item Some universally branching, or all existentially branching, premises are falsifiable,
	\item The variable conditions hold at $\gamma_0$.
\end{enumerate}
\end{lemma}
\begin{proof}
By induction on the length $h(\gamma_0)$ of the longest branch from $\gamma_0$ to a leaf sequent of $\tree{T}$.
\begin{description}
	\item[Base case:] $h(\gamma_0) = 0$. So $\gamma_0$ itself is an instance of $\IdRule$, $\FalseLeftRule$, $\TrueRightRule$, or $\ReturnRule$.
		\begin{description}
			\item[$\IdRule$, $\FalseLeftRule$, $\TrueRightRule$:] The conclusion of these rules is never falsifiable, so there is nothing to show.
			
			\item[$\ReturnRule$:] $ $

		The conclusion of the $\ReturnRule$ rule is $\Gamma \sequent \Delta$, and there is no premise. From the side condition of the $\ReturnRule$ rule, we know that no other rules are applicable to $\Gamma \sequent \Delta$. We will now show that $\Gamma \sequent \Delta$ is falsifiable, and that it obeys the variable conditions. 
			
			We create a model with a single world $w_0$, and for every atom $p$ in $\Gamma$, we let $\vartheta(w_0,p) = true$, and for every atom $q$ in $\Delta$, we let $\vartheta(w_0,q) = false$. Note that an atom cannot be both in $\Gamma$ and $\Delta$, since the $\IdRule$ rule in particular is not applicable to $\Gamma \sequent \Delta$.
			
			To show that $\Gamma \sequent \Delta$ is falsifiable at $w_0$, we need to show that $w_0 \Force \Gamma$ and $w_0 \Reject \Delta$. For every atom in $\Gamma$ and $\Delta$, the valuation ensures this. For every composite formula $\varphi$, we do a simple induction on its length. The fact that the $\ReturnRule$ rule is applied implies that no other rules are applicable, therefore the required subformula $\psi$ is already in $\Gamma$ or $\Delta$ as appropriate, and $\psi$ falls under the induction hypothesis.
			
			Thus we know that:
			\begin{description}
				\item[\rm{(i)}] $w_0 \Force \Gamma$ and
				\item[\rm{(ii)}] $w_0 \Reject \Delta$.
			\end{description}
			Then (i) and the persistence property of $\BiInt$ give us that $\forall w \in \mycal{W} :w_0 \mycal{R} w \MetaImp w \Force \Gamma$. Similarly, (ii) and the reverse persistence property of $\BiInt$ give us that $\forall w \in \mycal{W} .w \mycal{R} w_0 \MetaImp w \Reject \Delta$. Then the conclusion of the $\ReturnRule$ rule obeys the variable conditions:
			\begin{description}
				\item[$\Succ$-condition:] \textbf{S}uccessor condition \\
					$\exists \Sigma \in \{ \Gamma \} . \forall w \in \mycal{W} .w_0 \mycal{R} w \MetaImp w \Force \Sigma$
				\item[$\Pred$-condition:] \textbf{P}redecessor condition  \\
					$\exists \Pi \in \{ \Delta \} . \forall w \in \mycal{W} . w \mycal{R} w_0 \MetaImp w \Reject \Pi$				
			\end{description}			
		\end{description}
	\item[Induction step:] We assume that the lemma holds for all $\gamma_0$ with $h(\gamma_0) \leq k$, and show that it holds for all $\gamma_0$ with $h(\gamma_0) \leq k+1$. 
	
	Consider the rule application $\rho$ such that $\gamma_0$ is the conclusion of $\rho$. By the assumption of the lemma, we have that the conclusion $\gamma_0$ of $\rho$ is falsifiable at some $w_0$ in some model $\Model$. The only possibilities are that $\rho$ is a static or a special rule, or that it is a transitional rule:
	\begin{enumerate}
		\item $\rho$ is one of the static or special rules (universally branching). Then Lemma~\ref{staticRuleSoundness} tells us that some premise is falsifiable. We now need to show that the variable conditions hold at $\gamma_0$. There are two cases:
		\begin{description}
			\item[$\rho$ is unary:]
			The premise $\gamma_1$ of $\rho$ has $h(\gamma_1) \leq k$, therefore the induction hypothesis applies to $\gamma_1$. By Lemma~\ref{staticRuleSoundness} and the fact that $\gamma_0$ is falsifiable at $w_0$, we know that the premise $\gamma_1$ is falsifiable at $w_0$, so by the induction hypothesis we have that the variable conditions hold at $\gamma_1$. Since $\gamma_1$ has the same variables as $\gamma_0$, and since $\gamma_1$ is falsified by the same world $w_0$ as $\gamma_0$, we then know that $\gamma_0$ also obeys the variable conditions.
			\item[$\rho$ is $n$-ary with $n > 1$:] We show the case for $\Succ$; the case for $\Pred$ is symmetric. The premises $\gamma_1$ to $\gamma_n$ of $\rho$ each have $\gamma_i \leq k$, therefore the induction hypothesis applies to each $\gamma_i$. By Lemma~\ref{staticRuleSoundness} and the fact that $\gamma_0$ is falsifiable at $w_0$, we know that some $\gamma_m$ is falsifiable at $w_0$, too. Therefore the induction hypothesis tells us that the variable conditions hold at $\gamma_m$. That is, we know that:
	$$\exists \Sigma_m \in \Succ_m . \forall w \in \mycal{W} . w_0 \mycal{R} w \MetaImp w \Force \Sigma_m.$$
	To show that the conclusion $\gamma_0$ obeys the variable condition for $\Succ$, we need to show the following:
	$$\exists \Sigma \in \bigcup_1^n \Succ_i . \forall w \in \mycal{W} . w_0 \mycal{R} w \MetaImp w \Force \Sigma.$$
	Since $\Sigma_m \in \Succ_m$ and $\Succ_m \subseteq \bigcup_1^n \Succ_i$, we have $\Sigma_m \in \bigcup_1^n \Succ_i$ and thus  the variable conditions hold for $\Succ$ at the conclusion $\gamma_0$.
			\end{description}		
		\item $\rho$ is one of the transitional rules (existentially branching).
We show the case for the $\ImpRightRule$ rule, the case for the $\WeakImpLeftRule$ rule is symmetric:

{\scriptsize{\RuleDefImpRight}}\\

So suppose that the conclusion is falsifiable. Then we know from Definition \ref{falsifiability} that there exists a world $w_0$ such that:
		\begin{description}
			\item[\rm{(i)}] $w_0 \Force \Gamma$ and
			\item[\rm{(ii)}] $w_0 \Reject \Delta, \varphi \ExImp \psi$.
		\end{description}
	From the semantics of $\ExImp$ in $\BiInt$, (ii) implies that there exists a successor $w_1$ such that:
		\begin{description}
			\item[\rm{(iii)}] $w_0 \mycal{R} w_1$ and 		
			\item[\rm{(iv)}] $w_1 \Force \varphi$ and 
			\item[\rm{(v)}] $w_1 \NotForce \psi$.
		\end{description}	

		\begin{enumerate}
			\item\label{leftFalsifiable} To show that the left premise of the $\ImpRightRule$ rule is falsifiable, we need to show that there exists a world $w'$ such that this premise is falsifiable at $w'$. We let $w' = w_1$.
		
		Then items (i), (iv) and (v) give us that the left premise is falsifiable.
		
		Now, the left premise $\gamma_1$ is of distance $\leq k$ from the furthest leaf node of $\tree{T}$, therefore the induction hypothesis applies to $\gamma_1$. By the hypothesis assumption, since $\gamma_1$ is falsifiable at $w_1$, we have that the variable conditions hold at $\gamma_1$. In particular, the $\Pred$ condition holds, giving us:
\begin{equation}\label{predLeft}
	\exists \Pi \in \Pred_1 . \forall w \in \mycal{W} . w \mycal{R} w_1\MetaImp w \Reject \Pi
\end{equation}
		Now there are two cases: either the right premise was created, or it was not (and there is nothing to show). If it was created, then we need to show that it is falsifiable by exhibiting a world $w''$ such that the right premise is falsifiable at $w''$. We let $w'' = w_0$.
Then, since $w_0 \mycal{R} w_1$, we have $w_0 \Reject \Pi$ by \eqref{predLeft}. Since $\Pi \in \Pred_1$, then by Definition~\ref{extendedSemantics} we have that $w_0 \Reject \bigwedge \Pred_1$. Together with (i) and (ii), this means that the right premise is falsifiable at $w_0$.
		
		Moreover, the variable conditions hold at the right premise, since it also is falsifiable, and of distance $\leq k$ from the furthest leaf node of $\tree{T}$, so the induction hypothesis applies to it.

			\item We need to show that the variable conditions hold at the conclusion $\gamma_0$ of the $\ImpRightRule$ rule. We show the case for the variable $\Succ$; the case for $\Pred$ is symmetric. We need to show that:
\begin{equation}\label{succConcl}
	\exists \Sigma \in \Succ . \forall w \in \mycal{W} .w_0 \mycal{R} w \MetaImp w \Force \Sigma
\end{equation}
			Where $\varsTrans{\Succ}{\Succ_1}{\Succ_2}{\{ \Gamma \}}{\Pred_1}$
			
			Since we have shown that the variable conditions hold at the left premise, we know that in particular $\Pred_1 \ne \epsilon$. Therefore there are two cases: either the right premise was created, or it was not:
			\begin{itemize}
				\item If the right premise $\gamma_2$ was created, then we know that the variable conditions hold at $\gamma_2$, since $\gamma_2$ falls under the induction hypothesis. This gives us:
				$$\exists \Sigma_2 \in \Succ_2 . \forall w \in \mycal{W} . w_0 \mycal{R} w \MetaImp w \Force \Sigma_2$$
			Thus $\Succ := \Succ_2$ obeys \eqref{succConcl}.
				\item If the right premise was not created, then we need to show that $\{ \Gamma \}$ obeys the variable conditions at the conclusion. Now, we have $w_0 \Force \Gamma$ by (i), and then the persistence property tells us that $\forall w \in \mycal{W} .w_0 \mycal{R} w \MetaImp w \Force \Gamma$. Thus $\Succ := \{ \Gamma \}$ obeys \eqref{succConcl}.
			\end{itemize} 			
		\end{enumerate}		
		\end{enumerate}
	\end{description}		
\end{proof}

\subsection{Main Soundness Proof}

\begin{lemma}\label{presoundness} If $\Gamma \sequent \Delta$ is derivable then $\Gamma \sequent \Delta$ is not falsifiable.
\end{lemma}
\begin{proof}
By induction on the height $k$ of the derivation. 

\textbf{Base case:} For the base case, the height is 1. A derivation of height 1 can only be an instance of $\FalseLeftRule$, $\TrueRightRule$ or $\IdRule$. In each case, $\gamma$ is not falsifiable, as shown in cases 1 to 3 of Lemma \ref{staticRuleSoundness}.

\textbf{Inductive step:} We assume that if there is a derivation for $\gamma$ of height $\leq k$, then $\gamma$ is not falsifiable. We show that if there is a derivation for $\gamma$ of height $\leq k+1$, then $\gamma$ is not falsifiable. 

For a contradiction, suppose there is a derivation $\tree{T}$ for $\gamma$ of height $k+1$ and $\gamma$ is falsifiable. Consider the bottom-most rule application $\rho$ in $\tree{T}$, then $\gamma$ is the conclusion of $\rho$.

Then, by Definition~\ref{derivation}, since $\tree{T}$ is a derivation, then all universally branching premises, or some existentially branching premise of $\rho$ are rooted at derivations of height $\leq k$, so by the induction hypothesis, all universally branching premises are, or some existentially branching premise is not falsifiable. But since the conclusion $\gamma$ of $\rho$ is falsifiable by supposition, then by Lemma~\ref{globalSoundness}, some universally branching premise, or all existentially branching premises are falsifiable. Now we have a contradiction, therefore our assumption was wrong and $\gamma$ is not falsifiable.
\end{proof}

\begin{theorem}[Soundness] If $\Gamma \sequent \Delta$ is derivable, then $\Gamma \entails \Delta$.
\end{theorem}
\begin{proof}
By Lemma \ref{presoundness}, we have that $\Gamma \sequent \Delta$ is not falsifiable. Then by Lemma \ref{notFalsifiable}, we have $\Gamma \entails \Delta$.
\end{proof}

\section{Completeness}\label{completeness}

\subsection{Proof Outline}

We wish to prove:
\begin{itemize}
	\item[] if $\Gamma \entails \Delta$, then $\Gamma \sequent \Delta$ is derivable.
\end{itemize}
Instead, we prove the contrapositive:
\begin{itemize}
	\item[] if $\Gamma \sequent \Delta$ is not derivable, then there exists a counter-model for $\Gamma \entails \Delta$.
\end{itemize}

\noindent Our proof is based on a standard technique for proving completeness of tableau calculi: see \cite{gore1999}. We have adapted this technique to a two-sided sequent calculus with variables.

We assume that $\Gamma \sequent \Delta$ is not derivable, meaning that none of the $\GBiInt$-trees for $\Gamma \sequent \Delta$ is a derivation. Then we choose formulae from sequents found in possibly different $\GBiInt$-trees for $\Gamma \sequent \Delta$ in order to construct a counter-model for $\Gamma \entails \Delta$. The counter-model is constructed so that it contains a world $w_0$ such that $w_0 \Force \Gamma$ and $w_0 \Reject \Delta$, hence $\Gamma \entails \Delta$ does not hold.

\subsection{Saturated Sets}

\begin{definition}
Given a sequent $\Gamma \sequent \Delta$, we say that:
	\begin{itemize}
		\item $\Gamma \sequent \Delta$ is \textbf{consistent} if all of the following hold:
			\begin{enumerate}
				\item $\Bottom \not\in \Gamma$
				\item $\Top \not\in \Delta$
				\item $\Gamma \cap \Delta = \epsilon$				
			\end{enumerate}
		\item $\Gamma \sequent \Delta$ is \textbf{closed} with respect to a $\GBiInt$ rule $\rho$ if either:
			\begin{itemize}
				\item $\rho$ is not applicable to $\Gamma \sequent \Delta$, or
				\item Whenever $\Gamma \sequent \Delta$ matches the conclusion of an instance of $\rho$, then for some premise $\Gamma_1 \sequent \Delta_1$ of the instance of $\rho$, we have $\Gamma_1 \subseteq \Gamma$ and $\Delta_1 \subseteq \Delta$.
			\end{itemize}
		\item $\Gamma \sequent \Delta$ is \textbf{saturated} if it is consistent and closed with respect to the static rules of $\GBiInt$.
	\end{itemize}
\end{definition}

The following corollaries follow directly from the definition of consistent sequents.

\begin{corollary}\label{consNonId}
If $\Gamma \sequent \Delta$ is consistent, then none of the rules $\IdRule$, $\FalseLeftRule$, $\TrueRightRule$ is applicable to it.
\end{corollary}

\begin{corollary}\label{nonderIsConsistent}
If the sequent $$\SequentAny$$ is not derivable, then $\Gamma \sequent \Delta$ is consistent for all values of $\Succ$ and $\Pred$.
\end{corollary}

\begin{remark}\label{keepChoices}
  As usual, every sequent
  has a set of one or more ``saturations'' due to the branching of
  $\AndRightRule$, $\OrLeftRule$, etc., rules. The usual approach is
  to non-deterministically choose one of the non-derivable premises of
  each such rule. However, in the presence of the inverse relation, a
  branch that appears open may close once we return variables to a
  lower sequent. Therefore, we need to temporarily keep all the
  non-derivable premises, since we do not know which of the open
  branches will stay open when we return to a lower sequent.
\end{remark}

\begin{lemma}\label{saturationProcedure}
For each finite non-derivable sequent $\Gamma \sequent \Delta$, there is an effective procedure to construct a finite set $\zeta = \{ \alpha_1, \cdots, \alpha_n \}$ of finite saturated sequents, with $\Gamma \cup \Delta \subseteq LHS(\alpha_j) \cup RHS(\alpha_j) \subseteq sf(\Gamma) \cup sf(\Delta)$ for all $1 \leq j \leq n$.
\end{lemma}
\begin{proof}
Since $\Gamma \sequent \Delta$ is non-derivable, we know from Corollary~\ref{nonderIsConsistent} that $\Gamma \sequent \Delta$ is consistent. Then from Corollary~\ref{consNonId} we know that the $\IdRule$, $\FalseLeftRule$, $\TrueRightRule$ rules are not applicable to $\Gamma \sequent \Delta$. Let $\tree{T} = \Gamma \sequent \Delta$. While some static rule $\rho$ is applicable to a leaf of $\tree{T}$, extend $\tree{T}$ by applying $\rho$ to the leaf to obtain new leaves. Keep the non-derivable leaves only; by Corollary~\ref{nonderIsConsistent} they are consistent. By Theorem~\ref{termination}, the saturation process will eventually terminate; let $\zeta = \{ \alpha_1, \cdots, \alpha_n \}$ be the final leaves of $\tree{T}$. Since the formulae in each premise are always subformulae of the conclusion, we have that $LHS(\alpha_j) \cup RHS(\alpha_j) \subseteq sf(\Gamma) \cup sf(\Delta)$ for all $1 \leq j \leq n$.
\end{proof}

\subsection{Model Graphs and Satisfiability Lemma}

We shall use model graphs as an intermediate structure between $\GBiInt$-trees and $\BiInt$ models.

\begin{definition}\label{defModelGraph}
A model graph for a sequent $\Gamma \sequent \Delta$ is a finite $\BiInt$ frame $\pair{\mycal{W}}{\mycal{R}}$ such that all $w \in \mycal{W}$ are saturated sequents $\Gamma_w \sequent \Delta_w$ and all of the following hold:
	\begin{enumerate}
		\item $\Gamma \subseteq \Gamma_{w_0}$ and $\Delta \subseteq \Delta_{w_0}$ for some $w_0 \in \mycal{W}$, where $w_0 = \Gamma_{w_0} \sequent \Delta_{w_0}$;
		\item\label{p_impRight} if $\varphi \ExImp \psi \in \Delta_w$ then $\exists v \in \mycal{W}$ with $w \mycal{R} v$ and $\varphi \in \Gamma_{v}$ and $\psi \in \Delta_{v}$;
		\item\label{p_weakImpLeft} if $\varphi \WeakImp \psi \in \Gamma_w$ then $\exists v \in \mycal{W}$ with $v \mycal{R} w$ and $\varphi \in \Gamma_{v}$ and $\psi \in \Delta_{v}$;
		\item\label{p_impLeft} if $w \mycal{R} v$ and $\varphi \ExImp \psi \in \Gamma_w$ then $\psi \in \Gamma_{v}$ or $\varphi \in \Delta_{v}$;
		\item\label{p_weakImpRight} if $v \mycal{R} w$ and $\varphi \WeakImp \psi \in \Delta_w$ then $\psi \in \Gamma_{v}$ or $\varphi \in \Delta_{w'}$;
		\item\label{p_pers} if $w \mycal{R} v$ and $\varphi \in \Gamma_w$ then $\varphi \in \Gamma_{v}$;
		\item\label{p_revpers} if $v \mycal{R} w$ and $\varphi \in \Delta_w$ then $\varphi \in \Delta_{v}$.	
	\end{enumerate}
\end{definition}

\noindent We now show that given a model graph, we can use it to construct a $\BiInt$ model.

\begin{lemma}\label{modelGraphToModel}
If there exists a model graph $\pair{\mycal{W}}{\mycal{R}}$ for $\Gamma \sequent \Delta$, then there exists a $\BiInt$ model $\Model$ such that for some $w_0 \in \mycal{W}$, we have $w_0 \Force \Gamma$ and $w_0 \Reject \Delta$. We call $\mycal{M}$ the counter-model for $\Gamma \entails \Delta$.
\end{lemma}
\begin{proof}
Since we already have a $\BiInt$ frame $\pair{\mycal{W}}{\mycal{R}}$, we need to define a valuation $\vartheta$ in order to construct a $\BiInt$ model $\Model$:
\begin{enumerate}
	\item For every world $w \in \mycal{W}$ and every atom $p \in \Gamma_w$, let $\vartheta(w,p) = \true$.
	\item For every world $w \in \mycal{W}$ and every atom $q \in \Delta_w$, let $\vartheta(w,q) = \false$. 
\end{enumerate}

Then properties \ref{p_pers} and \ref{p_revpers} of Definition~\ref{defModelGraph} ensure persistence and reverse persistence respectively.

We now need to show that for every world $w \in \mycal{W}$, we have $w \Force \Gamma_w$ and $w \Reject \Delta_w$; we can do this by simple induction on  the length of the formulae in $\Gamma \sequent_w \Delta$.

Now let $w_0$ be the world in the model graph such that $\Gamma \subseteq \Gamma_{w_0}$ and $\Delta \subseteq \Delta_{w_0}$. Since our proof by induction has shown that for every world $w \in \mycal{W}$, we have $w \Force \Gamma_w$ and $w \Reject \Delta_w$, then in particular, we have that $w_0 \Force \Gamma_{w_0}$ and $w_0 \Reject \Delta_{w_0}$. Then, since we have that $\Gamma \subseteq \Gamma_{w_0}$ and $\Delta \subseteq \Delta_{w_0}$, we also have $w_0 \Force \Gamma$ and $w_0 \Reject \Delta$.
\end{proof}

\subsection{Main Completeness Proof}

\begin{figure}[t]
\textbf{Procedure} MGC \\
Input: sequent $\Gamma \sequent \Delta$ \\
Output: model graph $\pair{\mycal{W}^f}{\mycal{R}^f}$, variables $\Succ^f$ and $\Pred^f$
\begin{enumerate}
        \item Let $\zeta = \{ \alpha_1, \cdots, \alpha_n \}$ be the result of saturating $\Gamma \sequent \Delta$ using Lemma~\ref{saturationProcedure};
        \item\label{eachSaturatedVersion} For each $\alpha_i \in \zeta$ do 
                \begin{enumerate}
                \item Let $\pair{\mycal{W}_i}{\mycal{R}_i} = \pair{\{\alpha_i\}}{\{(\alpha_i, \alpha_i)\}}$; let $recompute := false$;
                \item\label{futureWorldConstruction} For each non-blocked $\varphi \Imp \psi \in \Delta_{\alpha_i}$ and while $recompute = false$ do
                \begin{enumerate}
                        \item\label{startSuccConstr} Apply $\ImpRightRule$ to $\varphi \Imp \psi$ and obtain a left premise $\pi_1 = \Gamma_{\alpha_i}, \varphi \sequent \psi$;
                        \item Let $\pair{\mycal{W}}{\mycal{R}}, \Succ, \Pred := MGC(\pi_1)$;
                        \item If $\exists \Pi_j \in \Pred . \Pi_j \subseteq \Delta_{\alpha_i}$ then
                        \begin{enumerate}
                                \item\label{chooseSuccessor} Let $u_j \in \mycal{W}_j$ be the root of the connected component $\mycal{W}_j$ from $\mycal{W}$;
                                \item\label{create} Let $G = \pair{\mycal{W}_j}{\mycal{R}_j}[j:=i]$; add $G$ to $\pair{\mycal{W}_i}{\mycal{R}_i}$, and put $\alpha_i \mycal{R}_i u_i$.
                        \end{enumerate}
                        \item\label{newInfo} else
                        \begin{enumerate}
                                \item\label{delete} Let $\pair{\mycal{W}_i}{\mycal{R}_i} = \pair{\epsilon}{\epsilon}$; let $recompute := true$; 
                                \item\label{addVars} Invoke the right premise of $\ImpRightRule$ to obtain $\pi_2 = \Gamma_{\alpha_i} \sequent \Delta_{\alpha_i}, \bigwedge \Pred$;
                                \item Apply $\SpecialRightRule$ to $\pi_2$ to obtain $m \geq 1$ non-derivable premises $\gamma_1, \cdots, \gamma_m$;
                                \item For each $\gamma_k$, $1 \leq k \leq m$, let $\pair{\mycal{W}_k}{\mycal{R}_k}, \Succ_k, \Pred_k  := MGC(\gamma_k)$;
                                \item\label{endSuccConstr} Let $\pair{\mycal{W}_i}{\mycal{R}_i} := \pair{\bigcup \mycal{W}_k}{\bigcup \mycal{R}_k}$, and $\Succ_{i} := \bigcup \Succ_{\gamma_k}$ and $\Pred_{i} := \bigcup \Pred_{\gamma_k}$;
                        \end{enumerate}                         
                \end{enumerate}
                \item\label{pastWorldConstruction} For each non-blocked $\varphi \WeakImp \psi \in \Gamma_{\alpha_i}$ and while $recompute = false$ do
                \begin{enumerate}
                        \item Perform a symmetric procedure to Steps~\ref{startSuccConstr} to \ref{endSuccConstr}.
                \end{enumerate}
                \item\label{finalVars} If $recompute = false$ then let $\Succ_{i} := \{\Gamma_{\alpha_i}\}$ and $\Pred_{i} := \{\Delta_{\alpha_i}\}$.
                \end{enumerate}
        \item Return $\pair{\bigcup \mycal{W}_i}{\bigcup \mycal{R}_i}, \bigcup \Succ_i, \bigcup \Pred_i$
\end{enumerate}
\caption{Model Graph Construction Procedure}
\label{algConstruction}
\end{figure}

We now show how to construct a model graph for
$\Gamma\hspace{-0.05cm}\sequent\hspace{-0.05cm}\Delta$ from a
consistent $\Gamma\hspace{-0.05cm}\sequent\hspace{-0.05cm}\Delta$.
Recall from Remark~\ref{keepChoices} that we need to keep a number of
independent versions of worlds because of the choices arising due to
disjunctive non-determinism. We do this by storing one or more
independent connected-components $\pair{\mycal{W}_1}{\mycal{R}_1},
\cdots, \pair{\mycal{W}_n}{\mycal{R}_n}$ in the constructed model
graph $\pair{\mycal{W}}{\mycal{R}}$, and the indices (sorts) of worlds
and relations tell us the connected-component of the graph to which
they belong.
We write $\pair{\mycal{W}_j}{\mycal{R}_j}[j:=i]$ to relabel the
connected component $\pair{\mycal{W}_j}{\mycal{R}_j}$ with sort $j$ to
a connected component $\pair{\mycal{W}_i}{\mycal{R}_i}$ with sort $i$.
Similarly, we also label each member of the variables $\Pred$ and
$\Succ$, so we can later extract the member 
with sort $i$, corresponding to the component
of $\pair{\mycal{W}}{\mycal{R}}$ with sort $i$. We write
$\mycal{R}$-neighbour to mean $\mycal{R}$-predecessor or
$\mycal{R}$-successor.

Our algorithm in Fig.~\ref{algConstruction} starts by saturating the
root world to obtain one or more saturated ``states''. For each
``state'' $\alpha_i$, it recursively creates all the
$\mycal{R}$-neighbours and saturates them, and so on.  If during the
construction of any $\mycal{R}$-neighbour, new information is returned
from the higher sequents (Step~\ref{newInfo}), then we delete the
entire subtree (connected component of sort $i$) rooted at $\alpha_i$,
and recreate $\alpha_i$ using the new information
(Step~\ref{addVars}). This re-creates all the
$\mycal{R}$-neighbours of $\alpha_i$. Otherwise, if none of the
$\mycal{R}$-neighbours of $\alpha_i$ return any new information, or
there are no $\mycal{R}$-neighbours for $\alpha_i$, then
Step~\ref{finalVars} instantiates the variables and returns from the
recursion. In the latter case, the ``state'' $\alpha_i$ already has
all the required information it can possibly receive from any
$\mycal{R}$-neighbours, thus $\alpha_i$ is final. Note the duality:
new information {\em from} a single $\mycal{R}$-neighbour means that
all of the members of a variable were new, while new information {\em
  at} a ``state'' $\alpha_i$ means that some $\mycal{R}$-neighbour
returned new information.

When we return from $MGC$, we form the union of the components of the
model graph and the variables from the different ``states'', so that
the caller of $MGC$ can extract the appropriate component at
Step~\ref{chooseSuccessor}.

\begin{remark}\label{proofVsCountermodel}
Note that while the counter-model construction procedure keeps the whole counter-model in memory, this procedure is only used to prove the completeness of $\GBiInt$. Our procedure for checking the validity of $\BiInt$ formulae (Fig.~\ref{strategyFig}) does not need the whole counter-model, and explores one branch at a time, as is usual for sequent/tableaux calculi.
\end{remark}

\begin{theorem}[Completeness]
$\GBiInt$ is complete: if $\Gamma \sequent \Delta$ is not derivable, then there exists a counter-model for $\Gamma \entails \Delta$.
\end{theorem}

\begin{proof}
Suppose $\Gamma \sequent \Delta$ is not derivable, then by Corollary~\ref{nonderIsConsistent} we have that $\Gamma \sequent \Delta$ is consistent. We construct a model graph for $\Gamma \sequent \Delta$ using the procedure given in Figure~\ref{algConstruction}, and obtain $\pair{\mycal{W}^f}{\mycal{R}^f}$. We let $\pair{\mycal{W}}{\mycal{R}}$ be any connected component of $\pair{\mycal{W}^f}{\mycal{R}^f}$. We now show that $\pair{\mycal{W}}{\mycal{R}}$ satisfies the properties of a model graph from Definition~\ref{defModelGraph}:
	\begin{enumerate}
		\item $\Gamma \subseteq \Gamma_{w_0}$
          and $\Delta \subseteq \Delta_{w_0}$ for some $w_0 \in
          \mycal{W}$: This holds because $w_0$ is one of the saturated
          sequents obtained from $\Gamma \sequent \Delta$.  Moreover,
          if we delete the original $w_0$ at Step~\ref{delete}, a
          final version of $w_0$ is created at Step~\ref{create} which
          is never deleted.
          
		\item\label{succExists} if $\varphi \ExImp
          \psi \in \Delta_w$ then $\exists v \in \mycal{W}$ with $w
          \mycal{R} v$ and $\varphi \in \Gamma_{v}$ and $\psi \in
          \Delta_{v}$: This holds because we have either created $v$
          using $\ImpRightRule$ at Step~\ref{create}, or had $w$
          fulfill the role of this successor by reflexivity if
          $\ImpRightRule$ was blocked.
          
		\item if $\varphi \WeakImp \psi \in \Gamma_w$ then there exists some $v \in \mycal{W}$ with $v \mycal{R} w$ and $\varphi \in \Gamma_{v}$ and $\psi \in \Delta_{v}$: \\
			By symmetry with property~\ref{succExists}.
			
		\item\label{impLeftAll} if $w
                  \mycal{R} v$ and $\varphi \ExImp \psi \in \Gamma_w$
                  then $\psi \in \Gamma_{v}$ or $\varphi \in
                  \Delta_{v}$: In our construction, there are three
                  ways of obtaining $w \mycal{R} v$, so we need to
                  show that for each case, the property holds. We
                  first show that $\varphi \ExImp \psi \in \Gamma_v$:
                        \begin{enumerate}
                                \item $v$ was created by applying $\ImpRightRule$ to $w$ on some $\alpha \ExImp \beta \in \Delta_{w}$. Then $\Gamma_v$ also contains $\varphi \ExImp \psi$.
                                \item $w$ was created by applying $\WeakImpLeftRule$ to some $\alpha \WeakImp \beta \in \Gamma_v$. Then, when the final version of $\Gamma_v$ was created, $\varphi \ExImp \psi \in \Gamma_w$ was added to the $\Succ$ variable at Step~\ref{finalVars}. There are two cases:
                                \begin{itemize}
                                        \item The right premise $\pi_2$ of $\WeakImpLeftRule$ was invoked at $v$. Then $\Succ$ was added to $\pi_2$ at $v$ by the symmetric process to Step~\ref{addVars}. Thus the updated $\Gamma_v$ also contains $\varphi \ExImp \psi$.
                                        \item The right premise of $\WeakImpLeftRule$ was not invoked at $v$. This means that $\exists \Sigma_j \in \Succ . \Sigma_j \subseteq \Gamma_v$, and the $j$-th version of $v$'s predecessor $w$ is chosen at the symmetric process to Step~\ref{chooseSuccessor}. But since Step~\ref{finalVars} at $w$ assigns $\Sigma_j  := \Gamma_w $, then we have $\Gamma_w \subseteq \Gamma_v$ and thus $\varphi \ExImp \psi \in \Gamma_v$.
                                \end{itemize}
                                \item $v = w$, and $w \mycal{R} w$ by reflexivity. Then $\Gamma_v = \Gamma_w$, so $\varphi \ExImp \psi \in \Gamma_v$.
                        \end{enumerate}         
                        In all cases, saturation for $v$ will then ensure that $\psi \in \Gamma_{v}$ or $\varphi \in \Delta_{v}$.
		\item if $v \mycal{R} w$ and $\varphi \WeakImp \psi \in \Delta_w$ then $\psi \in \Gamma_{v}$ or $\varphi \in \Delta_{v}$: \\
			By symmetry with property \ref{impLeftAll}.
		\item\label{forwardPersistence} if $w \mycal{R} v$ and $\varphi \in \Gamma_w$ then $\varphi \in \Gamma_{v}$: \\
			By similar argument to property \ref{impLeftAll}.
		\item if $v \mycal{R} w$ and $\varphi \in \Delta_w$ then $\varphi \in \Delta_{v}$:	\\	
			By symmetry with property \ref{forwardPersistence}.
	\end{enumerate}

\noindent We can obtain a counter-model for $\Gamma \entails \Delta$ from $\pair{\mycal{W}}{\mycal{R}}$ via Lemma \ref{modelGraphToModel}.
\end{proof}

\begin{definition}
A di-tree is a directed graph such that if the direction of the edges is ignored, it is a tree.
\end{definition}

\begin{theorem}
Every falsifiable $\BiInt$ sequent can be falsified by a model whose frame is a di-tree, consisting of reflexive points.
\end{theorem}
\begin{proof}
From Lemmas \ref{forwardOnlyFinite} and \ref{backwardOnlyFinite}, we know that the construction of new successors for $\varphi \ExImp \psi$ and predecessors for $\varphi \WeakImp \psi$ stops when either there are no rejected $\varphi \ExImp \psi$-formulae or forced $\varphi \WeakImp \psi$-formulae in the current world, or the current world already forces $\varphi$ and rejects $\psi$. In the latter case, the world itself fulfills the role of the successor or predecessor by reflexivity, and no new successors or predecessors are created.

The reason we are able to avoid proper cycles is the persistence and reverse persistence properties of $\BiInt$, used in the $\ImpRightRuleI$ and $\WeakImpLeftRuleI$ rules.

Consider the $\ExImp$ case. Every time some $\varphi \ExImp \psi$ appears on the RHS of a sequent $\Gamma \sequent \Delta, \varphi \ExImp \psi$, we first add $\psi$ to the RHS to obtain $\Gamma \sequent \Delta, \varphi \ExImp \psi, \psi$ using the $\ImpRightRuleI$ rule, since by reverse persistence the current world must reject everything that some successor world rejects. Now that $\psi$ is on the RHS, we need to apply the $\ImpRightRule$ rule to create the $\varphi \ExImp \psi$-successor $\Gamma, \varphi \sequent \psi$ only if $\varphi$ is not already on the LHS. For if $\varphi \in LHS$, then the successor $\Gamma \sequent \psi$ that fulfills $\varphi \ExImp \psi$ can be the current world itself. So there is no point creating it explicitly.
\end{proof}

\begin{corollary}
$\BiInt$ is characterised by finite rooted reflexive and transitive di-trees of reflexive points.
\end{corollary}

\section{Conclusions and Future Work}\label{sec:conclusion}

Our cut-free calculus for $\BiInt$ enjoys terminating backward
proof-search and is sound and complete w.r.t Kripke semantics.  A
simple Java implementation of $\GBiInt$ is available at
\url{http://users.rsise.anu.edu.au/~linda}.
The next step is to add a cut rule to $\GBiInt$, and prove cut
elimination syntactically. We are also extending our work to the modal logic $\SFive$, and the tense
logic $\KtSFour$. Our approach of existential branching and inter-premise communication bears some similarities to hypersequents of Pottinger and Avron~\cite{avron1996}. It would be interesting to investigate this correspondence further. From an automated deduction perspective, $\GBiInt$ is the first step towards an efficient decision procedure for $\BiInt$. The next task is to analyse the computational complexity of $\GBiInt$ and investigate which of the traditional optimisations for tableaux systems are still applicable in the intuitionistic case.

We would like to thank the anonymous reviewers for their suggestions.


\end{document}